\documentclass{aa}

\usepackage{graphicx}
\usepackage{amsmath}
\usepackage{natbib}
\usepackage{color}
\usepackage{enumitem}
\usepackage[T1]{fontenc}
\usepackage{subcaption}
\usepackage{placeins}
\usepackage{txfonts}
\usepackage{upgreek}
\newcommand{\mum}{$\upmu$m}

\definecolor{darkblue} {rgb}{0.0, 0.0, 0.7}
\usepackage[hyperindex=true, breaklinks=true,colorlinks=true,urlcolor={darkblue}, linkcolor={darkblue},  citecolor={darkblue}, bookmarks=true, bookmarksopen=true]{hyperref}

\newcommand{\Omegas}{\mathit{\Omega}}

\begin{document}

\title{Multi dust species inner rim in magnetized protoplanetary disks}

   \author{Mario Flock \inst{\ref{inst1}} \and Ond\v{r}ej Chrenko \inst{\ref{inst2}}  \and Takahiro Ueda \inst{\ref{inst1},\ref{inst3}} \and Myriam Benisty \inst{\ref{inst1}}
    \and Jozsef Varga \inst{\ref{inst5},\ref{inst6}} \and Roy van Boekel \inst{\ref{inst1}}}
   
   \institute{Max-Planck Institute for Astronomy (MPIA), K\"{o}nigstuhl 17, 69117 Heidelberg, Germany \email{flock@mpia.de} \label{inst1}
    \and
   Charles University, Fac Math \& Phys, Astronomical Institute, V Hole\v{s}ovi\v{c}k\'{a}ch 747/2, 180 00 Prague 8, Czech Republic \label{inst2}
    \and 
    Center for Astrophysics, Harvard \& Smithsonian, 60 Garden Street, M208, Cambridge, MA 02138, USA \label{inst3} 
    \and
   HUN-REN Research Centre for Astronomy and Earth Sciences, Konkoly Observatory, Konkoly-Thege Mikl\'{o}s \'{u}t 15-17, H-1121 Budapest, Hungary \label{inst5}
    \and 
    CSFK, MTA Centre of Excellence, Konkoly-Thege Mikl\'{o}s \'{u}t 15–17, 1121 Budapest, Hungary
    \label{inst6}
    }

\date{Received date /
Accepted date }
\abstract
{The inner regions of protoplanetary disks, within ten astronomical units (au), are where terrestrial planets are born. Developing multi-physics models of this environment is crucial for understanding how planets form.
By developing a new class of multi-dust radiative magnetized inner rim models and comparing them with recent near-infrared observational data, we can gain valuable insights into the conditions during planet formation.} {Our goal is twofold: to study the influence of highly refractory dust species on the inner rim shape and to determine how the magnetic field affects the inner disk structure. The resulting temperature and density structures are analyzed and compared to observations. The comparison focuses on a median SED of Herbig stars and interferometric constraints from the H, K, and N-band of three Herbig-type star-disk systems: \texttt{HD 100546}, \texttt{HD 163296}, and \texttt{HD 169142}.} {With the new models, we investigate 1) the influence of a large-scale magnetic field on the inner disk structure and 2) the effect of having the four most important dust species (corundum, iron, forsterite, and enstatite) shaping the rim, each with its sublimation temperatures.
Further, we improve our model by using frequency-dependent irradiation and the effect of accretion heating. With the \texttt{Optool} package, we obtain frequency-dependent opacities for each dust grain family and calculate the corresponding temperature-dependent Planck and Rosseland opacities. } {When multiple dust species are considered, the dust's sublimation front, i.e., the inner rim, becomes smoother and radially more extended. Strongly magnetized disks show a substantial increase in the emission flux between the L and N-bands. Our results show that weakly magnetized disk models with large-scale vertical magnetic fields $\leq 0.3$~ Gauss at 1~au best fit with NIR interferometric observations. Our model comparison supports the existence of moderate magnetic fields ($\beta \ge 10^4$), which could drive a magnetic wind in the inner disk. Our results show that multi-dust models, including magnetic fields, still lack NIR emission, especially in the H-band. Half-light radii derived from H-band emission by near-infrared interferometry indicate that the missing flux originates within the inner rim, where even corundum grains sublimate. One potential solution might be a heated gas disk or evaporating objects like planetesimals close to the star.}{}


\titlerunning{Multi-Dust Rim  }
\authorrunning{Flock et al.}

\maketitle

\section{Introduction}

How do planets form? With over 5000 detected exoplanetary systems, this is a crucial question of modern astrophysics. 
From the Kepler mission, focusing on planetary systems around F, G, K-type stars, we know that many rocky planets are located close to their host star, at orbital periods around ten days \citep{lis11,bat13,fab14,lis14}. Even more, multi-planetary systems show an 'inner edge', at which the occurrence rate of planets suddenly increases \citep{mul15,pet18,mul18} and which are often composed of 'super-Earth' planets in a compact configuration \citep{Wei23}. To understand the formation history of these compact super-Earth planets, we must understand the structure of the inner regions of the protoplanetary disk. 
In particular, the region where silicate material sublimates at temperatures around 1200 to 1400~K becomes of increased interest. Many important astrophysical characteristics come together here. At these locations and radially outward, the disk becomes optically thick to its thermal emission due to the condensation of the leading solid material components of iron (Fe), forsterite (Mg$_2$SiO$_4$), and enstatite (MgSiO$_3$) \citep{woi18}. Most of the near-infrared continuum flux is emitted at these locations, seen as a ring structure and often referenced to the dust rim \citep{dul10} located inside 1 au \citep{laz17,gra19}. 
Close to this location, the disk becomes ionized enough for the magneto-rotational instability to operate \citep{des15,flo17,thi19,les23,wil25}. This leads to a transition in the gas dynamics at the dust rim, having a very turbulent gas disk with turbulent velocities of several hundred meters per second \citep{flo17} and an equivalent $\alpha$-viscosity \citep{sha73} of about $\alpha \approx 0.1$ radially inward, while radially outward the level of turbulence drops several orders of magnitude \citep{tur14,les23}. Due to this transition in $\alpha$, we expect a steep rise in the gas surface density and the total disk pressure. This inner 'dead-zone' edge leads to an efficient concentration of solids, either in the form of pebbles which concentrate there at the local pressure maximum \citep{dzy10,flo17,ued19,flo19} or in the form of young inward migrating planets which can get trapped at these locations \citep{ida10,bol14,cha14,izi17,flo19,nes22,chr22,dra23}.

The best direct observational constraints for these inner planet formation regions come from either line or continuum emission in the near or mid-infrared. Observations of 51 Herbig AeBe objects by \citet{laz17} with the PIONIER instrument in the H-band (1.6 \mum) found evidence that the emission comes from the black body temperature of around 1800~K, higher than the typical sublimation temperature of the silicate grains. Observations by \citet{gra19} with the GRAVITY instrument in the K-band (2.2~\mum) are consistent with dust grains emitting around 1500~K. In a recent work by \citet{var24}, metallic iron as the innermost disk continuum component of HD~144432 yielded somewhat better fits to their VLTI data than amorphous carbon. A multiple-rim structure was postulated by \citet{mil16} using the Keck Interferometer in nuller mode \citep{col13}. As shown by \citet{var24}, equilibrium chemistry codes like \texttt{GGchem} \citep{woi18} can be used to explain the composition of the dust material in the inner disk regions.

On the theory side, new radiative hydrostatic models, including dust sublimation combined with hydrostatic equilibrium \citep{flo16}, recovered the detailed density and temperature regime and found distinct regions of a super-heated dust halo in front of the rim, a curved rim where silicates sublimate, and a shadowed region \citep{ued17,ued19}. \citet{vin21} demonstrated the importance of the radiation pressure force in accelerating dust grains of 1~$\mu$m in size at the inner rim radially outward. Further radiative hydrostatic models were presented by \citet{scho19} and \citet{scho21}, which included accretion heating and dust diffusion. Accretion heating is significant for the midplane temperature structure close to the inner rim, while dust diffusion affects the rim shape. In a recent work by \citet{chr24}, such radiative hydrostatic models were applied to the disk system \texttt{HD 163296} to compare in detail with near-infrared interferometric observations. Their model partially matched up to moderate baselines; however, they still had problems fitting the infrared excess in the H-band. The missing H-band flux problem was discovered three decades ago \citep{hil92}. Several researchers tried to solve this, using either a puffed rim wall \citep{dul01,nat01}, a dust halo \citep{vin06}, disk wind \citep{ban12}, magnetic pressure support \citet{tur14}, or the contribution of PAH's \citep{seo17}.

This work aims to investigate the role of multiple dust species and magnetic fields at the rim structure. { Highly refractory dust material like corundum can survive closer to the star.} We want to investigate if multiple dust species can lead to a second rim structure as postulated by \citet{mil16}. We consider four types of dust material: corundum, iron, forsterite, and enstatite, which constitute the main dust component close to the inner dust rim. We use \texttt{GGchem} \citep{woi18} to determine their sublimation curves. 

The second significant improvement is achieved by including the effect of a large-scale magnetic field by solving the magnetohydrostatic equations. We want to revisit the work by \citet{tur14}, which predicts increased H-band flux when including the magnetic pressure. In their work, they manipulated the density profile to mimic a puffed-up disk due to the magnetic pressure.  In our new class of models, we solve the magneto-hydrostatic equations directly for different large-scale magnetic field configurations and investigate their effect on the temperature and density structure. We also added the impact of accretion heating as it was implemented by \citet{scho19} and recently shown by \citet{chr24}. To set our nominal stellar and disk parameters, we focus on the typical Herbig star as it was used to explain the median spectral energy distribution (SED) by \citet{mul12} and which we also used in our previous work \citet{flo16} (model LS21). Additionally, we compare our results with a set of Herbig star-disk systems that share very similar stellar luminosity and for which we have constraints in the H, K, and N-band, namely \texttt{HD 163296}, \texttt{HD 169142}, and \texttt{HD 100546}. Even though these systems show a diverse disk density structure, having a similar stellar luminosity should enable a comparison to their inner rim structure with our models \citep{men15,laz17,gra19}.

The structure of the paper is as follows: In Section 2, we review the numerical method and the new developments. Section 3 presents the results of the new radiative magnetohydrostatic models of a typical Herbig star. In Section 4, we do the radiative transfer post-processing and compare our models with observations by calculating spectra and images at different wavelengths. Sections 5 and 6 follow with the discussion and the conclusion. 

\section{Method}
Our 2D axisymmetric model in the meridional plane follows the setup presented by \citet{flo16} and by \citet{flo19}. The following section reviews the main extensions and the steps to derive the radiation magnetohydrostatic equilibrium. The radiation magnetohydrostatic equilibrium is solved by iterating the radiative transfer, including irradiation, with the magnetostatic reconstruction. In the first step, the gas surface density profile is determined using
\begin{equation}
\Sigma(R) = \frac{\dot{M}}{3 \pi \nu_\mathrm{t}(R)}
\label{eq:sig_mdot}
\end{equation}
at the cylindrical radius $R$, assuming a constant radial mass accretion
rate $\dot{M}$, the viscosity 
\begin{equation}
\nu_\mathrm{t} = \frac{\alpha c_\mathrm{s}^2}{\Omegas} \, ,
\label{eq:nu_t}
\end{equation}
the sound speed $c_\mathrm{s}$, the disk rotation frequency $\Omegas=\sqrt{G
  M_*/R^3}$ {with the gravitational constant $G$, the stellar mass $M_*$} and the stress-to-pressure ratio
\begin{equation}
\alpha = \frac{1}{2} (\alpha_\mathrm{MRI} - \alpha_\mathrm{DZ} )  \left [ 1- \tanh{\left(
    \frac{T_\mathrm{MRI}-T}{T^\alpha_{\Delta}} \right )} \right ] + \alpha_\mathrm{DZ} \, ,
\label{eq:alpha}
\end{equation}
with the stress-to-pressure ratio in the active zone $\alpha_\mathrm{MRI}$ 
$(T>T_\mathrm{MRI})$, the stress-to-pressure ratio in the dead zone $\alpha_\mathrm{DZ}$, the ionization transition for the magneto-rotational instability (MRI) to operate $T_\mathrm{MRI}$, and the $\alpha$-transition temperature range $T^\alpha_{\Delta}$. { As in previous works, we use $T^\alpha_{\Delta} =100 \mathrm{K}$} to smooth the sudden transition of the $\alpha$ parameter. The magnetic field is set using the vector potential $\boldsymbol{A}$ and calculating $\boldsymbol{B} = \nabla \times \boldsymbol{A} $. The initial temperature field $T(r,\theta)$ is calculated using the
optically thin solution of a passive irradiated disk.
We then calculate the gas density $\rho(r,\theta)$
and the gas azimuthal velocity $v_\mathbf{\phi}(r,\theta)$ by solving
for magnetohydrostatic equilibrium in spherical geometry. Setting the time derivative to zero and assuming for the velocities $v_\mathbf{r}=v_\mathbf{\theta}=0$,  the equations in spherical coordinates are: 
\begin{eqnarray}
\nabla \cdot \left ( - B_\mathbf{r} \boldsymbol{B} \right ) + \frac{\partial P_\mathbf{t}}{\partial r} &=& - \rho \frac{\partial \Phi}{\partial
r} + \frac{\rho v^2_\mathbf{\phi} - B_\mathbf{\theta}^2 - B_\mathbf{\phi}^2}{r}  \label{eq:P_R} \\
\nabla \cdot \left ( - B_\mathbf{\theta} \boldsymbol{B} \right ) + \frac{1}{r} \frac{\partial P_\mathbf{t}}{\partial \theta} &=&
\frac{1}{\tan{\theta}}\frac{\rho v^2_\mathbf{\phi} - B_\mathbf{\phi}^2}{r}+\frac{B_\mathbf{r} B_\mathbf{\theta}}{r} \label{eq:P_T} \, 
\end{eqnarray}
where the gravitational potential is $\mathit{\Phi} = GM_*/r$, and $P_\mathbf{t}$ is the sum of magnetic pressure $\boldsymbol{B}^2/2$  and thermal pressure $P$ that relates to the
temperature $T$ through the ideal gas equation of state:
\begin{equation}
P= \frac{\rho k_\mathrm{B} T}{\mu_\mathrm{g} u},
\end{equation}
with the mean molecular weight $\mu_\mathrm{g}$, the Boltzmann constant $k_\mathrm{B}$ and the atomic mass unit $u$. The detailed equations used to derive each step's rotation velocity and density are presented in the Appendix \texttt{A}. We note that the magnetic field remains fixed during the iteration. To determine the magnetic field in each cell, we set the vector potential \begin{equation}
    \boldsymbol{A}=\begin{pmatrix} 0\\
    2 A_0 \\ A_0
    \end{pmatrix}
\end{equation} and calculate $\boldsymbol{B}=\nabla \times \boldsymbol{A}$ to obtain a vertical magnetic field with a $1/R$ profile with $R=r \sin{\theta}$ being the cylindrical radius and an azimuthal field which is twice as large at the midplane, having a $1/r$ profile: \begin{equation} 
    B_r = \frac{A_0 \cos{\theta}}{R}\\,
    B_\theta = \frac{A_0 \sin{\theta}}{R}\\,
    B_\phi = \frac{2 A_0}{r}
\end{equation}    
with $A_0$ being in units of [G cm]. In the following, we use the midplane value of $B_\theta$ to reference to the vertical magnetic field $B_z$ strength. Our reference model \texttt{M001} has a vertical field strength at 1~au of 10~mG. 

For a given density field, the radiation equilibrium is obtained as the steady-state solution to the following coupled set of
equations:

\begin{equation}
\label{eq:RAD1}
\frac{1}{\gamma-1}\partial_\mathrm{t} P = - \kappa_\mathrm{P}(T) \rho c (a_\mathrm{R} T^4 - E_\mathrm{R}) - \nabla \cdot F_* + Q_\mathrm{heat}
\end{equation}
\begin{equation}
\partial_\mathrm{t} E_\mathrm{R} - \nabla \cdot \left ( \frac{c \lambda}{\kappa_\mathrm{R}(T) \rho} \nabla
E_\mathrm{R} \right ) = \kappa_\mathrm{P}(T) \rho c (a_\mathrm{R} T^4-E_\mathrm{R})
\end{equation}

with the adiabatic index $\gamma$, the radiation energy E$_\mathrm{R}$,
the over frequency integrated irradiation flux $F_*$, the viscous heating $Q_\mathrm{heat}$,  the flux limiter $\lambda$ \citep{lev81}, the Rosseland and Planck opacity $\kappa_\mathrm{R}(T)$ and $ \kappa_\mathrm{P}(T)$, the radiation constant $a_\mathrm{R}=4 \sigma_\mathrm{b}/c$ with the Stefan-Boltzmann constant $\sigma_\mathrm{b}$, and $c$ the speed of light. {We choose for the mean molecular weight $\mu_\mathrm{g} = 2.353$ which represents a mixture between molecular hydrogen and helium. The adiabatic index is set to $\gamma = 1.42$ which is typically used for protoplanetary disks for this mixture \citep{dan13,bit13a,mar19}.}

\begin{figure}
  \resizebox{\hsize}{!}{\includegraphics{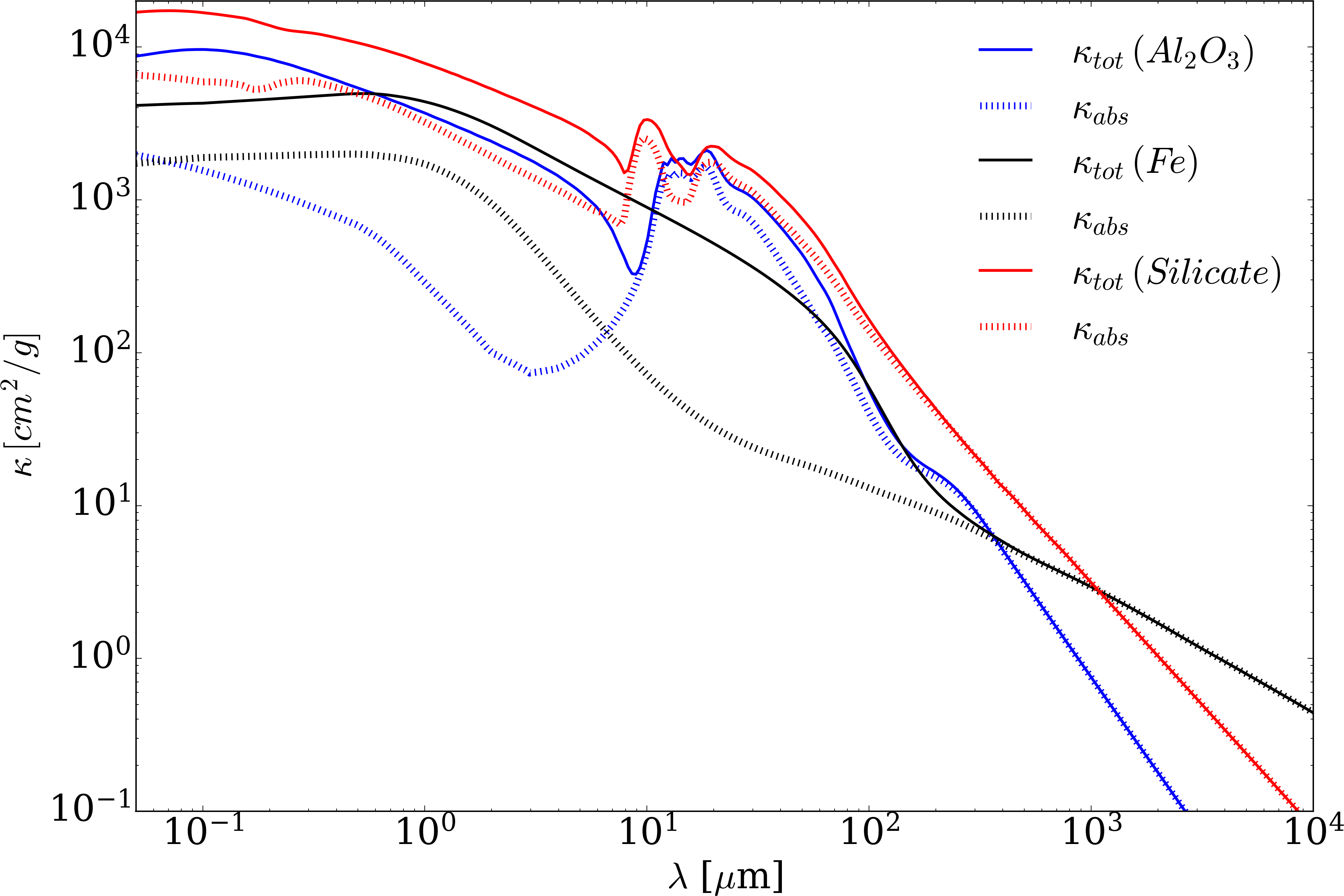}}
  \caption{Frequency-dependent opacity for the individual dust species. The solid line shows total opacity (absorption + scattering), and the dotted line shows absorption only.} 
\label{fig:opac}
\end{figure}

In this work, we include the effect of viscous heating, similar to what was done by \citet{scho19} using 
\begin{equation} Q_\mathrm{heat}= \rho \nu_\mathrm{t} [r \partial_\mathrm{r} \Omegas]^2. \end{equation} 
\citet{scho19} and \citet{chr24} reported difficulty obtaining a stable solution when including the $\rm \alpha$ temperature dependence. They both fixed the viscous heating to the value of $\alpha_\mathrm{DZ}$.
{Here, we first tried to include viscous heating based on Eq.~\ref{eq:alpha}, but we could not obtain a stable solution. Dynamical models are required to capture the viscous heating due to the dependence of $\alpha$ on the temperature itself \citep{cec24}.}  
{In this work, we also use} a constant $\alpha_\mathrm{DZ}$ everywhere in the disk. {Viscous heating due to} $\alpha_\mathrm{DZ}$ is a relevant heating source in the disk midplane regions beyond the inner rim, which become optically thick. { We refer to the work by \citet{cec24} which includes a fully dynamical heating and viscous evolution of a T Tauri star-disk model of the inner rim.}

\begin{figure}
\resizebox{\hsize}{!}{\includegraphics{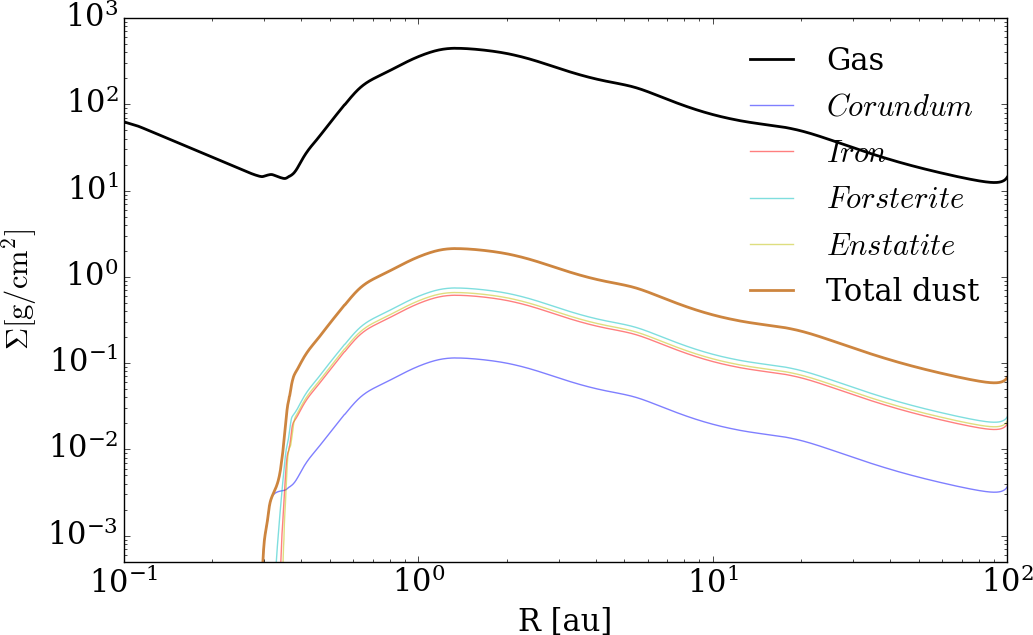}}
\resizebox{\hsize}{!}{\includegraphics{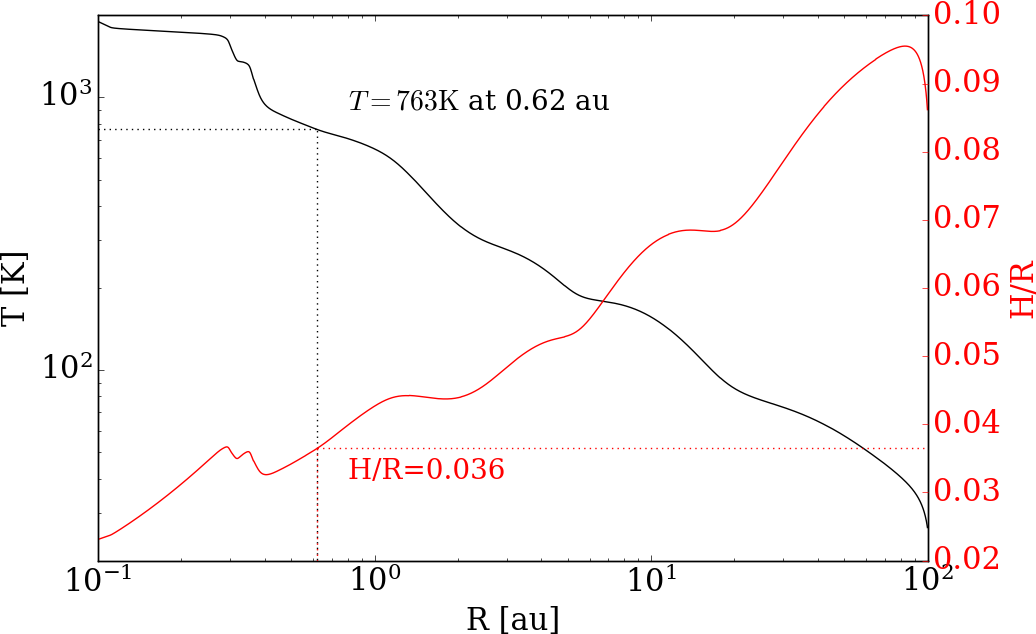}}
\caption{Top: Surface density profile of model \texttt{M001}. The gas is shown in black; the dust species are shown in color. Bottom: Midplane temperature profile (black) and H/R profile (red). The dotted line marks the position and corresponding temperature at the dust trap for which we would expect a substantial concentration of larger dust pebbles.} 
\label{fig:surf}
\end{figure}

\begin{table}
\caption{Overview of the dust species \label{tab:dust_spec} }
     \begin{tabular}{l|c|l}
       Dust species  &  mass ratio $f^s_0$ & $T^s_\mathrm{ev}$ ($\rho [\mathrm{g/cm^3}]$) \\
       \hline 
       $\mathrm{Al_2O_3}$ - Corundum  &  $2.571 \cdot 10^{-4}$ & $\rm 2237K \, \rho^{0.01578} $   \\ 
       $\mathrm{Fe}$ - Iron         &  $1.371 \cdot 10^{-3}$ & $\rm 2182K \, \rho^{0.02514} $  \\
       $\mathrm{Mg_2SiO_4}$ - Forsterite & $1.664 \cdot 10^{-3}$ & $\rm 1915K \, \rho^{0.01875} $ \\
       $\mathrm{MgSiO_3}$ -  Enstatite & $1.473 \cdot 10^{-3}$ & $\rm 1656K \, \rho^{0.01519} $ \\
\hline
    Total   & $4.766 \cdot 10^{-3}$  & 
    \end{tabular}
    \tablefoot{Dust species, their mass ratio, and their fitting function for sublimation. The mass ratio follows the standard solar parameter setup by \texttt{GGchem} \citep{woi18}. The fitting functions are derived in Appendix \texttt{B}.}
\end{table}

\begin{figure*}[t]
\resizebox{\hsize}{!}{\includegraphics{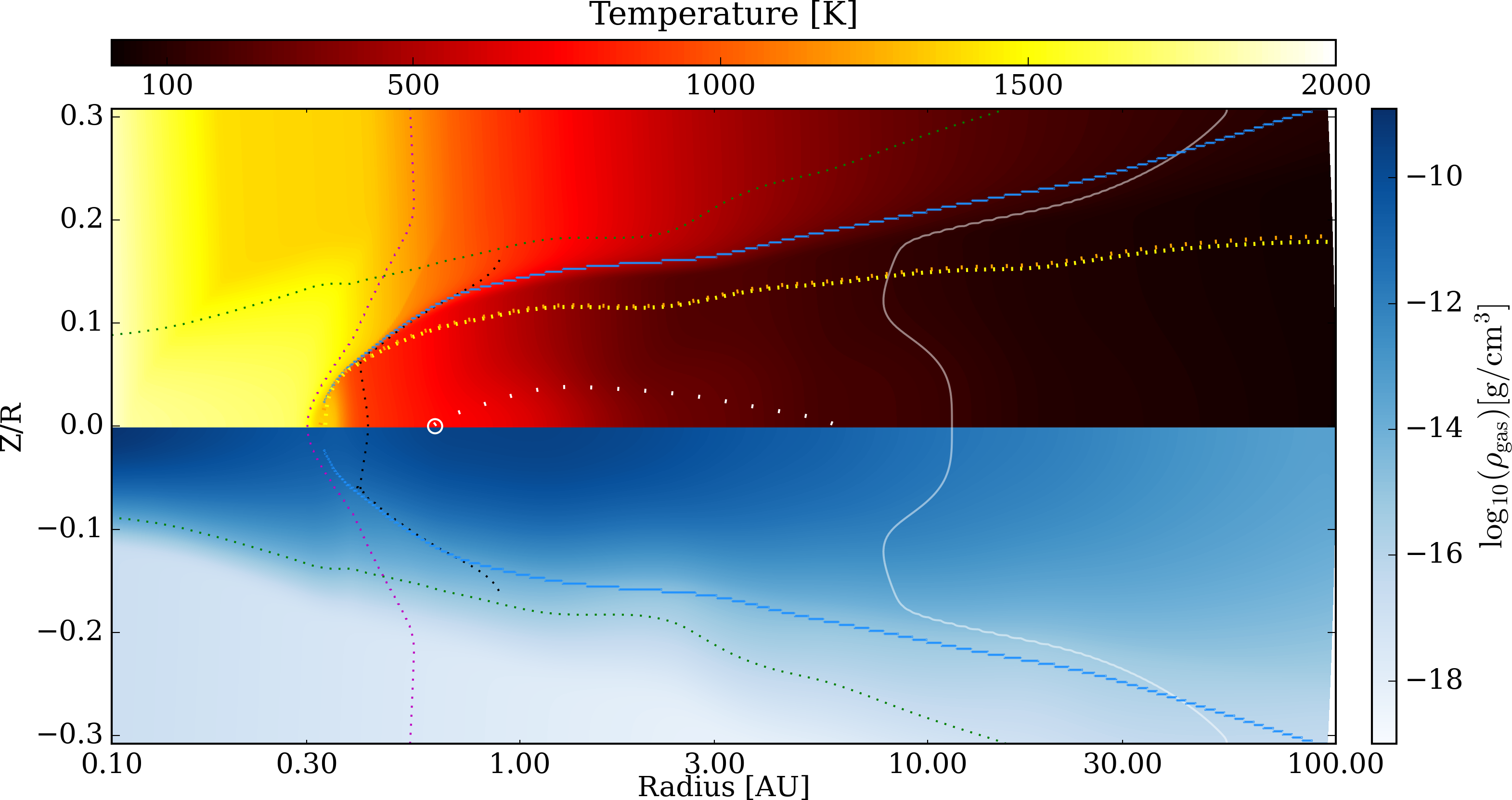}}
\caption{2D temperature and gas density profile of model \texttt{M001}. The green dotted line shows plasma beta unity. {The black dotted line shows the MRI transition (900~K contour line)}. The circle indicates the location of the pressure maxima where { inward drifting pebbles} can be trapped. The dotted magenta line shows the sublimation of the corundum. The white solid line shows the water snowline. The blue line shows the location of maximal irradiation heating corresponding to the $\tau_\mathrm{r} = 1$ line. {Yellow and orange small dashes show the $\tau_\mathrm{z} = 1$ line at 2.2 and 10 $\mu$m (K-band and N-band). White separated small dashes show the $\tau_\mathrm{z} = 1$ for 1~mm emission.}} 
\label{fig:inittemp}
\end{figure*}

\subsection{Multiple Dust Species}
\label{sec:dust_sub}
For this work, we use the tool \texttt{GGchem}  by \citet{woi18} to derive the composition and the sublimation temperature using four different dust species, $\mathrm{Al_2O_3}$  (corundum),   $\mathrm{Fe}$  (iron), $\mathrm{Mg_2SiO_4}$  (forsterite), and $\mathrm{MgSiO_3}$  (enstatite). Table~\ref{tab:dust_spec} summarizes the dust species, their dust-to-gas mass ratio, and their fitting functions for sublimation. The maximum value of the individual mass ratio is taken from \texttt{GGchem} and follows the values of our solar system. In this work, we chose these typical values that we expect to be present for the chosen Herbig star-disk model. We remind that changing the value of the total dust-to-gas mass ratio, which we assume here to describe the small grains ($\rm \leq 10 $ \mum), is similar to changing the total mass accretion rate, as both change the total vertical optical depth $\tau_\mathrm{z} = \kappa_\mathrm{P} \Sigma$. The effect of changing $\dot{M}$ was studied in our previous work; see \citet{flo16}. In this work, apart from the total dust-to-gas mass ratio, the ratio between the individual dust components is also taken from the standard solar model by \texttt{GGchem}. More details on the results of \texttt{GGchem} and how the fitting functions are derived can be found in Appendix \texttt{B}. To determine the sublimation temperature for each dust species, we use the fitting function $T^s_\mathrm{ev}$ from Table~\ref{tab:dust_spec}. Together with the maximum mass ratio $f^s_{0}$, we can determine in each step the abundance of the individual dust species using the following formula:

\begin{equation}
f^s_\mathrm{D2G}=\frac{f^s_0}{2} \left ( 1-\tanh \left ( \frac{T-T^s_\mathrm{ev}}{T_{\Delta}} \right )^3 \right ),
\label{eq:d2g}
\end{equation}
with $T_{\Delta}$ being the sublimation temperature range.

\subsection{Temperature transition $T_{\Delta}$ and $T^\alpha_{\Delta}$}
\label{Tdelta}
{ To stabilize the solution of the combined radiative transfer and hydrostatic models, we need to prevent steep gradients. These can occur in either the accretion stress to pressure ratio $\alpha$, see Eq.~\ref{eq:alpha}, or in the dust-to-gas mass ratio $f^s_\mathrm{D2G}$, see Eq.~\ref{eq:d2g}, to describe the dust density. A composition of only submicron silicate grains would sublimate over a short temperature range, usually shorter than what we could model with a grid capturing the whole disk. Solution to resolve the steep drop of the dust density at the silicate sublimation involves either multiple grain sizes \citep{kam09}, static mesh refinement of the dust rim \citep{vin12}, dust diffusion \citep{scho21}, or numerical smoothing \citep{flo16}. Models using a numerical smoothing with a single dust species without diffusion needed a smoothing of $T_{\Delta} = 100$~K, which was used by \citep{flo16,scho19,flo19,chr24}. Including multiple dust species has the advantage of reducing the value of $T_\Delta$ if needed, as the different sublimation temperatures of the dust species broaden the region over which the stellar irradiation is absorbed. 

Another transition happens for the accretion stress to pressure ratio due to the transition between a highly turbulent and a low-turbulent state of the disk. This transition at the inner dead-zone edge mainly depends on the dust microphysics, the temperature, the magnetic field, and the external radiation \citep{des15,wil25}. With a similar argument as before, previous works included a temperature transition to stabilize their models, while here models allowed a wider spread of values, from $T^\alpha_\Delta =100$~K used by \citet{flo16,scho19} to 70~K by \citet{chr24} to 25~K by \citet{flo19}. In this work, we decided to use for most of the models the same temperature transition for both the $\alpha$ transition and the dust sublimation and set $T^\alpha_\Delta = T_\Delta$. In our reference model, we use $T_\Delta = 100$~K while we also present models for a steeper dust sublimation temperature range $T_\Delta = 50$~K.}

\subsection{Opacity}
\label{sec:op}
The wavelength-dependent opacities are generated with the \texttt{Optool}\footnote{ \href{https://github.com/cdominik/optool}{\texttt{Optool}}, see Appendix \texttt{B} for detailed command parameter} package for the different dust family species; see Table~\ref{tab:optool}. The frequency-dependent opacities, absorption, and total opacity per gram of dust are shown in Fig.\ref{fig:opac}. The frequency-dependent absorption opacity is used to determine the irradiation heating\footnote{In this model, we follow the frequency-dependent irradiation as by \citet{flo13} (section 2.1) using 64 frequency bins.}. The irradiation flux is determined as
\begin{equation}
    F_*=\int_{\Omegas_\mathrm{s}}\int_{\nu_\mathrm{f}} B(\nu_\mathrm{f},T_*) \left ( \frac{R_*}{r} \right )^2 e^{-\tau_\mathrm{r}(\nu_\mathrm{f},r)} \Omegas_\mathrm{s} d\nu_\mathrm{f} d\Omegas_\mathrm{s}
\end{equation}
 with the Planck function $B(\nu_\mathrm{f},T_*)$, the solid angle $\Omegas_s$, the surface temperature of the star $T_*$, the radius of the star $R_*$, the frequency $\nu_\mathrm{f}$, and the radial optical depth 
 \begin{equation}
     \tau_\mathrm{r}(\nu_\mathrm{f},r) = \int_{R_*}^{r}  \kappa(\nu_\mathrm{f}) \rho dr.
 \end{equation}
For the region between $R_*$ and the first cell in our domain, we take the gas opacity into account as the dust is fully sublimated. To determine the total absorption opacity in each cell, we sum the individual dust components  $\kappa(\lambda)=\sum_s f^s_\mathrm{D2G} \kappa^s_\mathrm{abs}(\lambda) + \kappa_\mathrm{gas}$ with $\lambda=c/\nu_\mathrm{f}$. The gas opacity is set to the low value of $\kappa_\mathrm{gas}=10^{-6} \rm cm^2/g$ and is constant along temperature and wavelength.

We generated mean Planck $\kappa_\mathrm{P}(T)$ and Rosseland's opacities $\kappa_R(T)$ from the wavelength-dependent opacities for the three dust family species, dependent on the local temperature. We note that we use the same silicate opacity for Forsterite and Enstatite. We include the total extinction opacity to determine the Rosseland opacity. To determine the total Planck and Rosseland opacity in each cell, we sum again over each dust component using $\kappa_\mathrm{P}(T) = \sum_s f^s_\mathrm{D2G} \kappa_\mathrm{P}(T) + \kappa_\mathrm{gas}$. {More information on the opacity calculation can be found in the Appendix \texttt{B}.}

\subsection{Model setup}

We highlight again that compared to our previous models by \citet{flo19}, we improve our models by three significant changes: 1) including frequency-dependent irradiation, 2) adding multiple dust species at different sublimation temperatures, and 3) adding large-scale magnetic fields. The first two main changes help to resolve the dust rim { from a numerical point of view}, and we do not need additional smoothing to avoid a sudden increase in $\Delta \tau_{r}$. This is because the location for radial optical depth unity $\tau_\mathrm{r}(\lambda)=1$ differs for each wavelength, which broadens the layer where most of the irradiation heating is taking place. Also, the various dust components sublimate at different radii, smoothing out the radial profile of the total $f_\mathrm{D2G}$ and so avoiding a sudden jump of dust density as is the case using a single dust species. We tested Eq.~\ref{eq:d2g} for different resolutions in Appendix \texttt{D} to show convergence. Further discussion and alternative solutions for a single dust component of $f_\mathrm{D2G}$ can be found in \citet{flo16} (Eq. 10), \citet{flo19} (Eq. 9), or \citet{scho21} (Eq. 4), which uses dust diffusion to smooth out the rim.

The final disk model is reached by iteratively solving for the radiation equilibrium solution and magnetohydrostatic equilibrium, including frequency-dependent irradiation from the star. Another difference to our previous model by \citet{flo16} and \citet{flo19} is that we use a cold boundary\footnote{{ A cold boundary sets the radiation energy density in the ghost cells to a small value. Therefore, the cooling efficiency is increased, and the resulting temperatures are more similar to temperatures obtained by pure Monte Carlo radiative transfer simulations of a passive irradiated disk, like it is done using \texttt{RADMC3D} (mctherm).}} when determining the value of the radiation energy density in the ghost cells at the meridional boundary. However, we note that it remains difficult to argue what the correct boundary condition is ( read more in the Appendix by \citet{chr24}). Our new models use a fixed temperature of 10~K within the ghost cells. Table~\ref {tab:setup} summarizes the model input parameters.

\begin{table}
\caption{Setup parameters for the disk models. \label{tab:setup}}
\begin{tabular}{l|ll}
\hline
Stellar parameters & $T_*=8500\, \mathrm{K}$, $R_*=2.12\, {\rm R_\sun}$\\ 
                  & $M_*=2.0\, {M_\sun}$\\
Opacity for dust species $s$ & $\kappa^s(\lambda)$ see Fig.~\ref{fig:opac}\\
        & $\kappa_\mathrm{gas} = 10^{-6}\, \mathrm{cm^2/g}$\\
$N_\mathrm{r} \times N_\mathrm{\theta} $ & 2304 x 216 \\ 
$R_\mathrm{in}-R_\mathrm{out} : Z/R $ & 0.1--100~au : $\sim \pm 0.3$\\ 
Ionization transition & $T_\mathrm{MRI} = 900~\mathrm{K}$\\
{ Sublimation temperature range} & $T_\Delta =$ 100~K or 50~K\\
  $\alpha$ viscosity &  $\alpha_\mathrm{MRI} = 10^{-1}$ for $T > T_\mathrm{MRI}$\\
                       & $\alpha_\mathrm{DZ} = 10^{-3}$ for $T \leq T_\mathrm{MRI}$\\
$\alpha$-transition temperature range & $T^\alpha_\Delta =$ 100~K\\
  Dust-to-gas mass ratio & see Table ~\ref{tab:dust_spec} \\
Mass accretion rate        & $\dot{M}=1.0 \times 10^{-8} M_\sun/{\rm yr}$\\
\hline
\end{tabular}
\end{table}

\begin{table*}[ht]
\caption{Overview of the magnetized disk models. \label{tab:info}}
\begin{tabular}{llllllll}
\vspace{3mm}\\
Model & $B_z$ [G] & $\beta_z$ at 1~au & $T_\Delta$ & $R^\mathrm{in}_{\rm rim}$ & $R^{\rm out}_{\rm rim}$ & $\Delta R_{\rm rim}$ & $\tau_\mathrm{z}=1$ at $R^{\rm out}_{\rm rim}$ [rad] \\
\hline
\texttt{M001} & 0.01 & $2.29 \cdot 10^6$ & 100~K & 0.33~au & 0.52~au & 0.19~au & 0.085\\
\texttt{M01} & 0.1 & $2.29 \cdot 10^4$ & 100~K & 0.33~au & 0.52~au & 0.19~au & 0.085\\
\texttt{M03} & 0.3 & $2.51 \cdot 10^3$ & 100~K & 0.33~au & 0.52~au & 0.19~au & 0.085\\
\hline
\texttt{M001\_S} & 0.01 & $2.29 \cdot 10^6$ & 100~K & 0.42~au & 0.54~au & 0.12~au & 0.082\\
\texttt{M001\_S\_TC} & 0.01 & $2.29 \cdot 10^6$ & 100~K & 0.26~au & 0.40~au & 0.14~au & 0.073\\
\hline
\texttt{MZ\_S} & $10^{-4}$ & $1.52 \cdot 10^{10}$ & 100~K & 0.42~au & 0.54~au & 0.12~au & 0.082\\
\texttt{MZ} & $10^{-4}$ & $1.52 \cdot 10^{10}$ & 100~K & 0.33~au & 0.52~au & 0.19~au & 0.085\\
\texttt{MZ\_TD50} & $10^{-4}$ & $1.52 \cdot 10^{10}$ & 50~K & 0.36~au & 0.56~au & 0.20~au & 0.093\\
\end{tabular}
\tablefoot{From left to right: model name, the midplane vertical magnetic field strength at 1~au, and the plasma beta of the vertical magnetic field at the midplane at 1~au.}
\end{table*}

\begin{figure*}[t]
\resizebox{\hsize}{!}{\includegraphics{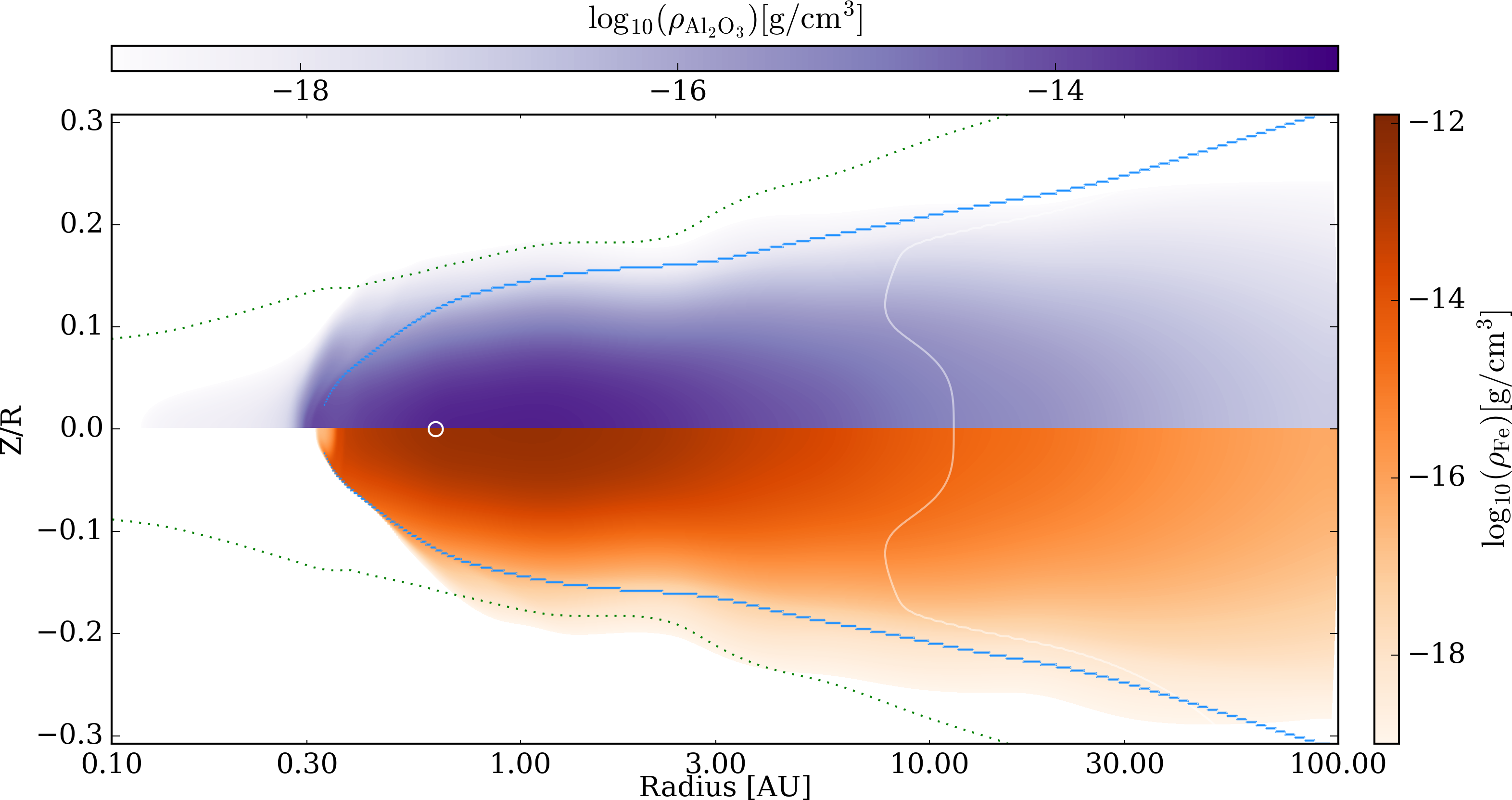}}
\caption{2D dust density profile of model \texttt{M001} for corundum (upper half) and iron (lower half). {A small amount of corundum survives as a hot optically thin halo in front of the inner rim ($\tau_\mathrm{r}=1$ blue contour line).}} 
\label{fig:initdust}
\end{figure*}

We follow previously established parameters of the radiative hydrodynamical models
by \citet{flo16}, model LS21, which should represent a class of Herbig-type stars used to derive a median SED by \citet{mul12}. For the ionization transition $T_\mathrm{MRI}$, we assume 900~K, which is set by thermionic emission of dust grains \citep{des15}. The accretion rate is constant over radius at $10^{-8} M_\sun/{\rm yr}$. Compared to our previous models, we extend the domain to 100~au to capture the whole near and mid-infrared emitting region of protoplanetary disks. The turbulent stress-to-pressure ratio is set to $\alpha_\mathrm{MRI}=0.1$ and $\alpha_\mathrm{DZ}=0.001$ {which follows the results obtained from 3D radiation MHD simulations of the inner rim with a vertical magnetic field \citep{flo17}}.
To generate the initial conditions, we use a resolution of 2304 cells in radius with logarithmically increasing cell size and 216 cells in the meridional direction with uniform spacing for a domain extent from 0.1 to 100~au in radius and $\Delta \theta =0.6\, \mathrm{rad}$ ($\sim 34^\mathrm{o}$). The inner domain boundary was chosen to reach hot enough temperatures so that the corundum dust component fully sublimates. For the different models, we vary the magnetic field strength, starting with a low value of $B_\mathrm{z}= 10 \, \mathrm{mG}$ at 1 au, corresponding to a plasma beta ($\beta = 2P/B^2$) of $2.26 \times 10^6$ at the midplane after the radiation magnetohydrostatic equilibrium is reached for model \texttt{M001}. We increase the magnetic field strength successively until reaching a field strength of $B_\mathrm{z} = 0.3~\mathrm{G}$ in the model \texttt{M03} corresponding to a plasma beta of $2.51 \times 10^3$. Following magnetic wind-driven accretion models, \citet{bai09} predict a minimum magnetic field of $0.1~\mathrm{Gauss}$ to sustain an accretion rate of $10^{-8} M_\sun/{\rm yr}$ in the dead-zone, which would correspond to model \texttt{M01}. A similar magnetic field strength, corresponding to a plasma beta value of around $\beta=10^4$, was also found by \citet{les21} to be sufficient to sustain such accretion rates by a magnetic-driven wind. Even stronger magnetized models are presented in the Appendix \texttt{E}, reaching the maximum $B_z = 3~ \mathrm{Gauss}$ in model \texttt{M3}.

\begin{figure}
\resizebox{\hsize}{!}{\includegraphics{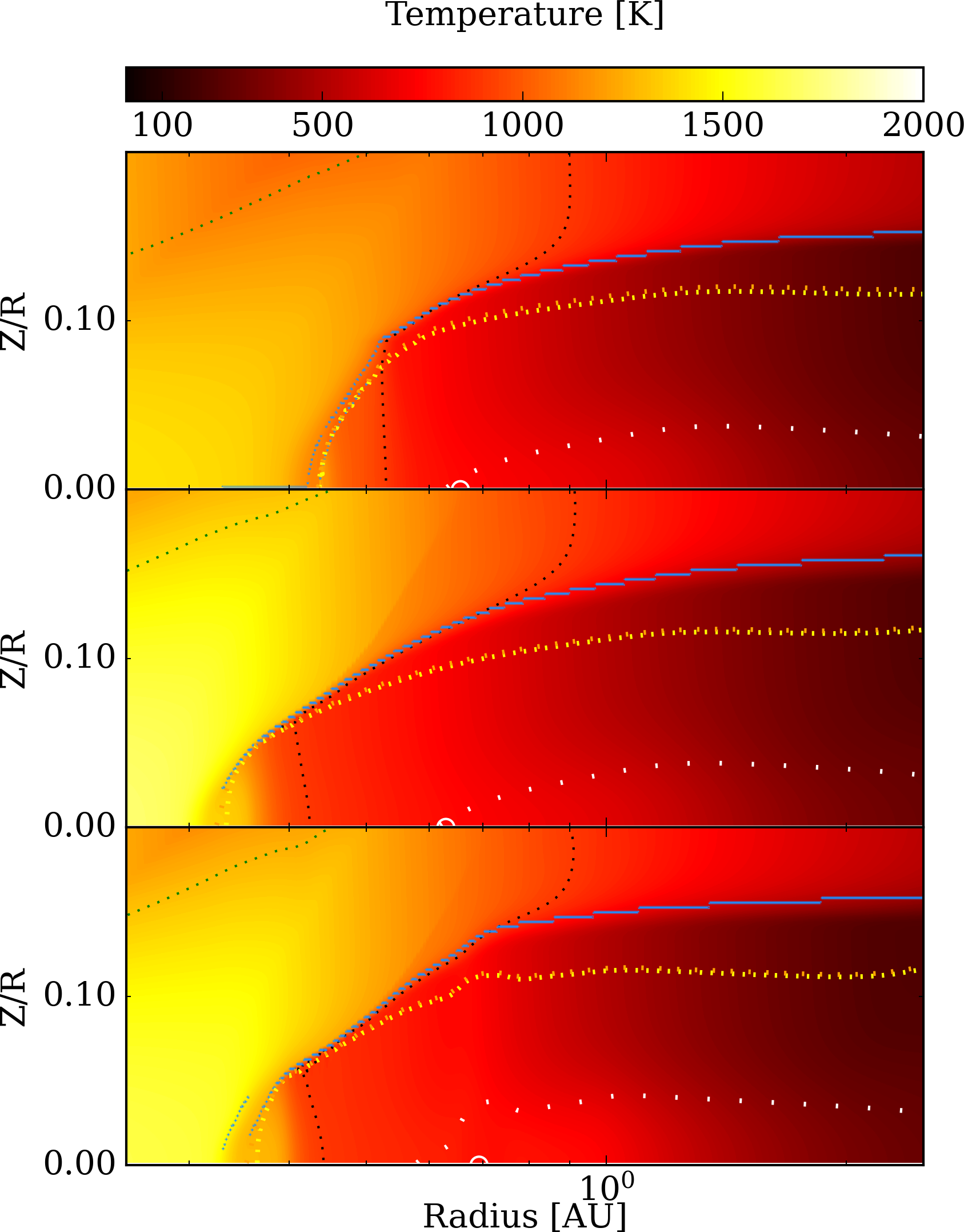}}
\caption{{ Models with very low magnetic field, zooming onto the temperature structure at the inner dust rim.  From top to bottom: Model \texttt{MZ\_S} with a single dust species; Model \texttt{MZ} and model \texttt{MZ\_TD50} both with multiple dust species. Model \texttt{MZ\_TD50} uses a steeper transition to change from $\alpha_{\rm MRI}$ to $\alpha_{\rm DZ}$. Annotations follow Fig.~\ref{fig:inittemp}.}} 
\label{fig:shape}
\end{figure}

\section{Results}
We first show the results of our reference model \texttt{M001}. The radial surface density profiles of the gas and dust of model \texttt{M001} are shown in Fig.~\ref{fig:surf}, top. As shown in our previous works, the rise in the gas surface density becomes visible due to decreased accretion stress in the dead zone \citep{flo16,flo19}. Due to thermionic emission, the disk remains long enough ionized for the MRI to operate \citep{des15}, and the surface density rises beyond, but still close to, the dust sublimation front, around 0.34 au. 

 As shown in our previous models, due to the rise in surface density, there is a clear pressure maximum and a corresponding trap of solid material located at around 0.62~au, indicated with a black dotted line in Fig.~\ref{fig:surf}, bottom. The temperature reaches 763~K at this dust trap with an H/R of 0.036. Fig.~\ref{fig:surf}, top, also shows that the corundum grains survive the closest to the star, however, contributing only a tiny fraction to the total dust surface density. Iron, forsterite, and enstatite show nearly the same sublimation region at 0.35~au. Only the corundum sublimates inward at 0.3~au.

The radial profiles of midplane temperature and aspect ratio H/R are shown in Fig.~\ref{fig:surf}, bottom. The first substantial temperature drop happens at around 0.34~au due to the sudden dust condensation. Then follows a shallow drop in temperature due to the curved rim, which changes the flaring and, therefore, the grazing angle of incoming irradiation at the disk surface between 0.4 and 1~au. The next steeper temperature drops happen between 1 and 3~au. Here, the accretion heating efficiency drops, which leads to a steeper drop in the temperature. There are two shadowed regions at 10 and 20~au, which can be seen in the H/R profile. In a shadowed region, the H/R remains constant over radius. In our models, these shadowed locations converge and do not propagate inward between the final iteration steps. { In general, disk shadowing events were reported in previous works as well \citep{wu21,ued21,fuk22,kut24}, and more dynamical simulations are needed to study their effect on the disk structure in the future}.\\
\begin{figure*}[t]
\resizebox{\hsize}{!}{\includegraphics{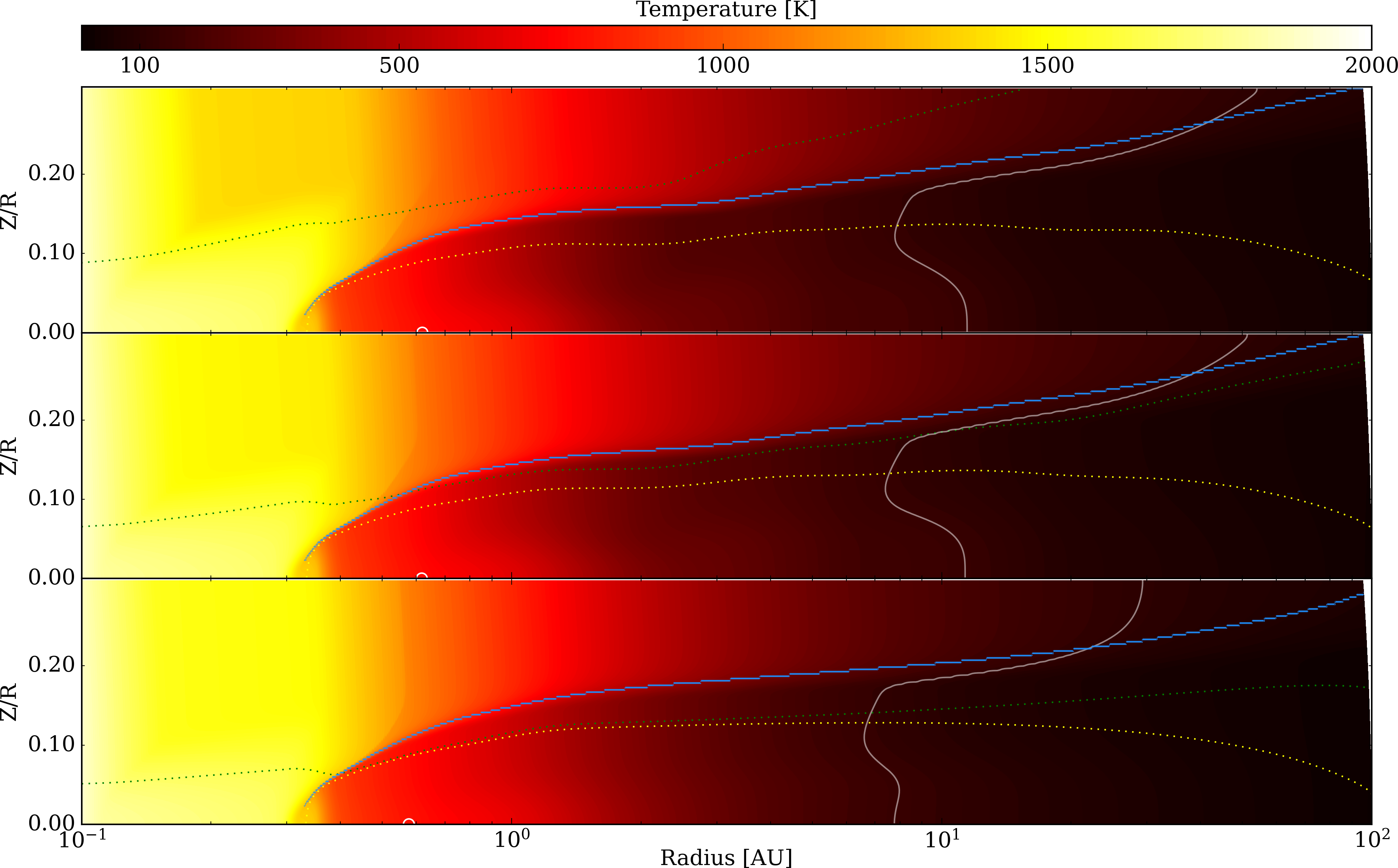}}
\caption{2D temperature profile with increasing magnetization from top to bottom showing models: \texttt{M001},\texttt{M01} and \texttt{M03}. The green dotted line shows the plasma beta unity line. The blue line shows the location of maximal irradiation heating corresponding to the $\tau_\mathrm{r} = 1$ line. Using the local Planck Opacity, the yellow dotted line shows the $\tau_\mathrm{z} = 1$ line. The white solid line shows the water snowline.}  
\label{fig:tempmag}
\end{figure*}
Fig.~\ref{fig:inittemp} shows the 2D temperature and density structure of model \texttt{M001}. The temperature structure and the rim look broader and smoother compared to our previous model \texttt{LS21} by \citet{flo16}. We observe a hot dust halo in front of the rim. There is a round rim shape at which the temperature suddenly drops. Another main difference is the contribution of accretion heating, which results in the midplane being warmer than the layers below the irradiation absorption between 1~au and 3~au. There is a small shadowed region between 1 and 3~au and a flared disk beyond 3~au. { For the reference model \texttt{M001}, the vertical magnetic field strength is 10~mG at 1~au. In the next subsection, we describe in more detail the influence of the magnetic field on the inner rim structure.}
The green dotted line shows the location where magnetic pressure equals thermal pressure. For model \texttt{M001}, the plasma beta unity line lies above the irradiation surface and in the optical thin layer. Due to the accretion heating, the snow line curves vertically and extends beyond 10~au in the midplane region, while in the disk atmosphere, it reaches further inward. To illustrate the snow line, we follow our previous routine by \citet{flo19} using the formula by \citet{oka11}, which determines the water vapor mol fraction
\begin{equation}
  X_\mathrm{H_2O}= \left \{
      \begin{array}{ll} 
        1.2 \times 10^{-3} (1.2 \times P \le P_{\rm sat} (T) ),\\
        P_{\rm sat}(T)/P  (1.2 \times P \ge P_{\rm sat} (T),\\
      \end{array}
      \right.
\end{equation}
with the saturated vapor pressure
\begin{equation}
P_{\rm sat} = e^{-6070 \rm K/T+30.86} \rm dyn\, cm^{-2}.
\end{equation}
We then determine the position where half of the water is in the form of ice by
$\rm x_{ice}=0.5$ with
\begin{equation}
x_\mathrm{ice}= 1 - X_\mathrm{H_2O}/(1.2 \times 10^{-3}).
\end{equation}

Fig.~\ref{fig:inittemp}, bottom, presents the corresponding 2D gas density profile. The gas density is highest at the midplane at around 0.1 and 1 au, following the surface density profile. The H/R variations connected to the shadowing also become visible in gas density variations, especially in the upper layers. This can be seen best by looking at the pathway of the plasma beta unity line (green dotted line), which shows a jump between 1 and 3~au. 
In Fig.~\ref{fig:inittemp}, we also show the sublimation of the corundum with the magenta dotted line. The steep temperature drop happens slightly radially outward compared to the dotted magenta line, which is because corundum condenses at small abundances that do not suffice to make the disk optically thick to stellar irradiation. The corundum thus contributes to the hot, optically thin emission. { Fig.~\ref{fig:inittemp} also shows where the maximum of irradiation heating occurs (blue line), which is similar to the stellar spectrum weighted $\tau_\mathrm{r}=1$ line. At the same time, the vertical rise of the line illustrates the grazing angle of the irradiation. The yellow and orange dotted lines trace the $\tau_\mathrm{z}=1$ at 2.2 and 10 $\rm \mu m$, which are almost tracing the identical disk surface and are both located below the $\tau_\mathrm{r}=1$ line. To trace the midplane region, longer wavelengths are required, see the white separated dots, which trace the $\tau_\mathrm{z}=1$ at 1 mm.}

To disentangle the spatial distribution of various grain species, we show the corresponding 2D profile of corundum and iron dust grains in Fig.~\ref{fig:initdust}. The figure shows that a small amount of corundum survives in front of the inner rim due to the higher sublimation temperature than iron and the silicate group. The corundum grains constitute the heated dust halo before the rim. In contrast, the iron density shows a sudden and sharp drop at the rim. The sublimation temperature of iron is very close to enstatite and forsterite, see Appendix \texttt{B}. %In the midplane, the dust volume densities reach values up to $\rm 10^{-12} g/cm^3$. 
In both Fig.~\ref{fig:inittemp} and Fig.~\ref{fig:initdust}, we indicate the maximum pressure with a white circle at the midplane.
In the following two sections, we introduce models with nearly no magnetization ($\beta = 1.52 
\cdot 10^{10}$), with and without multi-grain species, and with different levels of $T_\Delta$ to better distinguish their effect on the rim shape.

\subsection{Rim shape and width}
To study the effect of multiple dust species on the rim shape, we created two models with only one dust component, with and nearly without magnetic fields, model \texttt{M001\_S} and model \texttt{MZ\_S}. These models use the same total dust-to-gas mass ratio, given in Table~\ref{tab:dust_spec}, and assume only a single silicate species represented by forsterite. The results are summarized in Fig.~\ref{fig:shape} and Table~\ref{tab:info}. We measure the size and extent of the inner rim by determining the innermost location in the disk where $\tau_\mathrm{z}=1$ and the outermost location where the temperature drops below 800~K at $\tau_\mathrm{z}=1$ \citep{flo16}. Further, we determine the height of this location. Both pure silicate models \texttt{M001\_S} and \texttt{MZ\_S} show a compact inner rim, starting at 0.42~au with a 0.12~au extent reaching the outer rim. The different levels of magnetization show no effect on the position or extent of the rim, which is expected as here the thermal pressure dominates by many orders of magnitude. Adding multiple dust species makes the rim rounder and smoother. For model \texttt{MZ}, the inner rim moves inward, starting at 0.33~au with an extent of 0.19~au. 
The normalized ring width for the multi dust species model $\Delta R_{\rm rim}/R^{\rm in}_{\rm rim}$ is 0.58 compared to 0.28 for the pure silicate models \texttt{MZ\_S} and \texttt{M001\_S}. A larger normalized ring width seemed to be what is observed when looking at K-band observations with the VLTI (see Fig.~3, left in \citet{gra19}).

As reported in section~\ref{Tdelta}, the temperature range $T_\Delta$ which controls the transition between the MRI turbulent inner disk and the more laminar outer disk at $T_\mathrm{MRI}$ can influence the rim structure. To distinguish the effect of this transition region, we created model \texttt{MZ\_TD50}, which uses a transition temperature of $T_\Delta=50$~K. The effect can be seen in Fig.~\ref{fig:shape}, bottom. The rim surface becomes slightly steeper compared to the reference models with $\Delta T=100$~K, shifting the rim a little bit outward from 0.33~au to 0.36~au and $\Delta R_{\rm rim}$ from 0.19~au to 0.2~au. At the same time, the height of the rim is increased here as well, from $\Delta \theta=0.085$ above the midplane to $\Delta \theta=0.093$.

\subsection{Effect of magnetization}
To study the effect of different magnetic fields, we show the 2D temperature profiles of the models with higher magnetization. Fig.~\ref{fig:tempmag} shows the different models with increasing magnetization from top to bottom. \texttt{M001} and \texttt{M01} are shown in the first and second row. The magnetization increases, which becomes visible in the shift of the plasma beta unity line closer to the midplane; see green dotted line. { The overall shape and extent of the inner rim are similar for all three levels of magnetizations, see Table~\ref{tab:info}.} For higher magnetization, the disk becomes more puffed up, and the higher density leads to an extended heated dust halo inside the inner rim (compare temperature at $R=0.2$~au and $Z/R=0.2$ for different magnetized models in Fig.~\ref{fig:tempmag}). Furthermore, higher magnetization reduces the grazing angle in the outer disk (beyond 3~au). Therefore, we observe that the ice line moves inwards, both at the midplane and the upper layers. At the midplane, the reduced density due to the larger vertical height leads to a reduction of the viscous heating. In the upper layers, two effects reduce the temperature and move the ice line inwards. First, the overall higher density leads to a higher optical depth, and so a reduction of the temperature in the disk atmosphere (above the $\tau_\mathrm{r}=1$ line), and second, the grazing angle becomes reduced and so reduces the efficiency of irradiation absorption which can be seen by looking at the rise in the height of the blue line at around 10~au in Fig.~\ref{fig:tempmag}. This reduction of the grazing angle is visible beyond 4~au in our models, radially inwards, the magnetic fields lead to an increase of the grazing angle due to the magnetic pressure support. We show the maximum heating by irradiation with the blue contour line, corresponding to the location of $\tau_\mathrm{r}=1$. We determine the height at 1~au of this location to $Z=0.144$~au for model \texttt{M001} and \texttt{M01}. For model \texttt{M03}, we measure an increase to $Z=0.146$~au. { Our results show that the influence of moderate magnetic field strength becomes larger for the outer disk regions beyond 1~au, affecting the grazing angle of stellar starlight and so the disk flaring. The influence on the disk regions inside 1~au and in particular around the dust rim and the emitting region of the K-band emission remains small, as here the thermal pressure is much larger than the magnetic pressure.} 

\begin{table}
\caption{Setup parameter for RADMC-3D \label{tab:radmc3d}}
\begin{tabular}{l|p{3.5cm}}
\hline
dustopac.inp & using 4 dust species (see Table~\ref{tab:dust_spec})\\
dustkappa\_*.inp & \texttt{Optool} opacity files (see Table~\ref{tab:optool})\\
camera\_wavelength\_micron.inp & spectrum output using 64 wavelength points (log scale between 1 $\mu$m and 1~mm)\\
nphot                        & $10^7$\\
scattering\_mode\_max          & 1 (isotropic scattering)\\
itempdecoup                  & 1 (dust temperature decouple)\\
inclination                  & $45^\circ$ (for images only, otherwise zero)\\
\hline
\end{tabular}
\tablefoot{For generating images at 2.2 and 4.8 ${\rm \mu m}$, we use npix=1000 points and zoom into -5 to 5~au.}
\end{table}

\begin{table*}[ht]
\caption{Comparison to NIR interferometry constraints of observed systems\label{tab:comp}}
\begin{tabular}{l|l|l|ll|llll|l}
Model/Star & $\rm log_{10}(\dot{M})$ & $L_*$ & $f_d^H$ & $f_d^K$ & $a_\mathrm{H}$ [au] & $a_K$ [au] & $a_L$ [au] & $a_N$ [au] & $F_N$ [Jy]\\
\hline
\texttt{M001} & $-8$  & 21 & 0.28 & 0.46 & 0.31 & 0.34 & 0.42 & 0.79 & 25.7  \\
\texttt{M03} & $-8$ & 21 & 0.3 &  0.51 & 0.34 & 0.42 & 0.49 & 0.93 & 36.2 \\
\hline
\texttt{M001\_S\_TC} & $-8$  & 21 & 0.40 & 0.62 & 0.24 & 0.24 & 0.25 & 0.79 & 22.0 \\
\hline
\texttt{HD 163296} & $ -6.79^{+0.15}_{-0.16} $ & 15.8 & 0.76 $\pm$ 0.01 & 0.81 $\pm$ 0.01 & 0.22 $\pm$ 0.05 & 0.3 $\pm$ 0.01 & 0.33 $\pm 0.01$ &  0.82 $\pm$ 0.05 & 16.3 $\pm$ 0.6\\
\texttt{HD 169142} & $<-8.7$ & 20.4 & 0.17 $\pm$ 0.01 & 0.40 $\pm$ 0.05 & 0.11 $\pm$ 0.02 \footref{fn:hband} & 0.33 $\pm$ 0.07 & - & 9.5 $\pm$ 2.3 & 0.92 $\pm$ 0.3\\
\texttt{HD 100546} & $\approx -7$ & 22.5 & 0.42 $\pm$ 0.02 & 0.63 $\pm$ 0.01 & 0.26 $\pm$ 0.01 & 0.28 $\pm$ 0.01 & - & 21.3 $\pm$ 1.1 & 69.1 $\pm$ 1.5\\
\hline
\end{tabular}
\tablefoot{From left to right listing: model name, accretion rate in units of $M_\sun/{\rm yr}$, stellar luminosity, fraction of disk emission in H and K-band, half-light radius for H, K, L, and N-band in units of au, N-band flux normalized at 100~pc in units of Jy. 
{Accretion rates were measured using X-Shooter observations and fitting the H$\alpha$ emission for \texttt{HD 163296} \citep{wic20}, \texttt{HD 169142} \citep{gra07} and \texttt{HD 100546} \citep{mend15}.} H-band information on the disk selection is retrieved by \citet{laz17} Table B.2 and \href{http://cdsarc.u-strasbg.fr/viz-bin/qcat?J/A+A/599/A85}{CDS Link}. K-band from \citet{gra19} Table~3. L-band information was taken for \texttt{HD 163296} from \citet{var21}. N-band from \citet{men15} Table~2. Gaia DR3 distances are assumed for \texttt{HD 169142} (114.9 pc), \texttt{HD 163296} (101 pc), and \texttt{HD 100546} (110~pc).}
\end{table*}

\begin{figure}
\resizebox{\hsize}{!}{\includegraphics{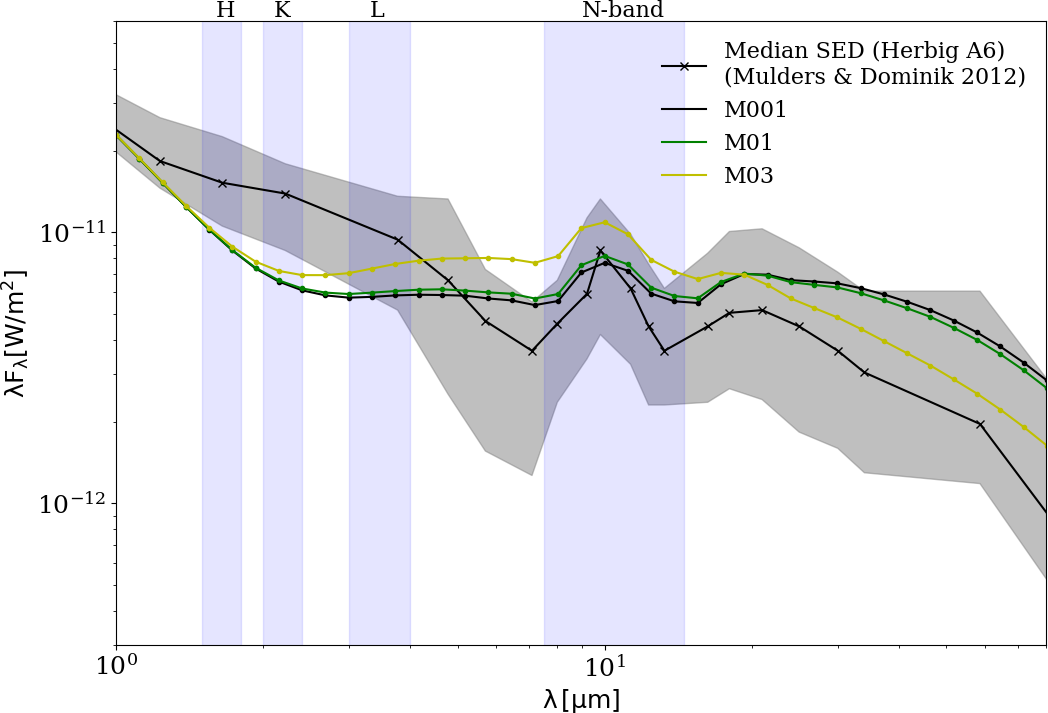}}
\caption{SED for the different magnetized models from \texttt{M001} to \texttt{M03} calculated at 64 wavelengths (dots). The black solid line (crosses) and grey shaded color show the median SED and one-sigma error constructed by \citet{mul12}. All values shown here were normalized to a distance of 100~pc.} 
\label{fig:sed}
\end{figure}

\section{Radiative transfer post-processing}
In the following section, we post-process the dataset with the Monte-Carlo radiative transfer code RADMC-3D\footnote{\href{https://www.ita.uni-heidelberg.de/~dullemond/software/radmc-3d/}{RADMC-3D 2.0}} by \citet{dul12}. The input parameters are summarized in Table~\ref{tab:radmc3d}. We first recalculate the dust temperature with RADMC-3D with the {\it mctherm} command using the 2D density of all four dust species. We ignore the gas opacity for this step. Here, we allow the individual dust species' temperatures to decouple and assume isotropic scattering. In the second step, we calculate the images and the spectrum at different wavelengths.

\subsection{SED comparison}
In this subsection, we determine the SED for the individual magnetizations and compare it with the median SED, constructed by \citet{mul12}. 
The median SED comes with an upper and lower quartile information, telling us the range around the median where 50\% of all flux values lie. We can use it to understand if our model represents a typical Herbig system or rather an outlier. Similar median SEDs were constructed for T Tauri stars \citep{Fur06}. The spectrum for the different magnetized models is shown in Fig.~\ref{fig:sed}. The low and moderate magnetized models \texttt{M001} and \texttt{M01} show very similar SEDs even though model \texttt{M01} has a midplane plasma beta with a factor 100 higher with $\rm \beta_z = 2.29 \times 10^{-4}$ at 1~au. As shown in Fig.~\ref{fig:tempmag}, this can be explained by looking at the position of the $\rm \beta=1$ line, which lies above or close to the $\rm \tau_\mathrm{r}=1$ line. 
Here, the magnetization only affects the uppermost, low-density, and optically thin layers of the disk and so does not affect the temperature structure or the overall shape of the SED. Even though we did not fine-tune the model, the profile of the low- and moderate-magnetization models compares well with the median SED, mainly staying within the one-sigma deviations. The 10-$\rm \mu m$ feature in the median SED appears stronger than for our models and choice of opacity (comparing the flux ratio between 6 and 10 $\rm \mu m$). { This might indicate that our models are too optically thick, especially in the dead zone region, just outside the dust rim.} Further, the H and K-bands still show missing flux, similar to our previous work \citep{flo16}. The hot corundum grains that survive higher temperatures contribute to hot optically thin emission; however, the SED profile shows this contribution is insufficient. At the H-band, the median flux is 1.7 times higher than the low magnetization models \texttt{M001} and \texttt{M01}. At the N-band at the peak of the silicate feature, the total flux of the low magnetized models matches very well with the median SED. At the same time, in the far infrared at 58 \mum, the low magnetization models overshoot the emission flux by around two. For the higher magnetized model \texttt{M03} with a midplane $\beta_\mathrm{z}=2.51 \times 10^3$ at 1~au, the SED shows a clear increase of emission flux between the K and N-bands. The increase in emission flux at the K-band brings model \texttt{M03} SED closer to the median SED. However, the model emission flux between 6 to 10 \mum \ moves away from the median value. Overall, we think model \texttt{M03} is less comparable to the median SED. The magnetic pressure increases the disk surface, affecting the stellar absorption layer, as shown in the previous section. The results of model \texttt{M03} should be interpreted with care, as we would expect strong magnetic torques at the magnetic wind base for model \texttt{M03}, leading to a reduction of the surface density and so a change in the overall structure.

\subsection{H-band comparison}
First, we focus on the H-band observations with PIONIER/VLTI of Herbig type stars by \citet{laz17}. As we have seen from the comparison to the median SED, our models still miss flux in the H-band. The observations described in \citet{laz17} measured the fraction of H-band emission flux, which comes from the disk, and the radial position, where half of the emission is coming from. The half-light radii are the results of parametric fitting, and \citet{chr24} showed that this result can be used as a robust observational constraint. We chose observations of three Herbig stars with similar stellar luminosities to our model. The results are summarized in Table~\ref{tab:comp}. We note that we used the distances from the GAIA DR3 release. The errors in the table are taken directly from the individual measurements of the different publications. What becomes directly visible is that the half-light radius in the H-band of the three objects lies systematically inward compared to our models. \texttt{HD 163296} and \texttt{HD 100546} show similar positions, with $0.22 \pm 0.05$~au and $0.26 \pm 0.01$~au. { \texttt{HD 169142} has a half-light radius in the H-band of $0.11 \pm 0.02$~au\footnote{We note that 0.11~au corresponds to 1~mas, close to the resolution limit of the VLTI PIONEER instrument.\label{fn:hband}}, while an independent measurement with the CHARA long-baseline interferometer shows a best fit of $0.21 \pm 0.01$~au \citep{Set18}.} For our fiducial model \texttt{M001}, we determined the H-band half-light radius to $0.31$~au, while the higher magnetization model shows even a slight increase to $0.34$~au for model \texttt{M03}. This comparison shows that most of the missing H-band flux should come from inside the silicate dust rim. Although we include corundum grains in our models, which survive close to the star at higher temperatures, they only slightly increase the H-band flux. The discussion section will outline some possible physical processes that could cause the additional compact emission.
Further, we can compare the disk H-band emission fraction to the stellar contribution. For the models, we determine the H-band emission fraction from the disk by using the emission flux from our model domain starting at 0.1~au. From the VLTI observations, this is done by fitting three components to the visibility profile: an unresolved stellar component that manifests as a plateau at longer baselines, resolved disk emission at shorter baselines, and extended emission. { Extended emission in H-band could be caused by scattered starlight, however the previous observations of these three objects report strong variations}\footnote{For H-band: \texttt{HD 163296} has an extended emission fraction $f_{ext}$ of zero, while \texttt{HD 100546} has $f_{ext}=0.11$ and \texttt{HD 169142} has $f_{ext}=0.08$.}. In our models, roughly one-third of the emission in the H-band comes from the disk. We determine a fraction of $0.28$ of disk H-band emission for the model \texttt{M001}, increasing to $0.3$ for model \texttt{M03}. For the selection of Herbig systems, the values are diverse, from $0.76 \pm 0.01$ for \texttt{HD 163296}, to $0.42 \pm 0.02$ in \texttt{HD 100546}, until $0.17 \pm 0.01$ in \texttt{HD 169142}. This also shows that at least the two specific systems follow the trend we saw for the median SED comparison: the overall flux in the H-band coming from the disk is higher than in our models. One might think no infrared excess is needed to describe the H-band flux for the system \texttt{HD 169142} due to the low fraction of disk emission in the H-band. However, previous works also needed to add hot dust emission at 1800~K to fit the H-band emission \citep{tsc21}.

\subsection{K-band comparison} 
For the K-band comparison, we focus on the observations and data obtained with GRAVITY/VLTI by \citet{gra19}. All three objects show very similar K-band half-light radii, with $0.33 \pm 0.07$~au for \texttt{HD 169142}, over $0.3 \pm 0.01$~au for \texttt{HD 163296} to $0.28 \pm 0.01$~au for \texttt{HD 100546}. These results match our fiducial model \texttt{M001}, which has a half-light radius of $0.34$~au. The higher magnetized models show values of $0.42$~au for model \texttt{M03}, less comparable to the observations. { Here, the larger radius is an effect due to higher K-band emission at larger radii (see section \ref{sec:intprof}).} The spread in the fraction of disk emission between the three observed objects is smaller in the K-band compared to the H-band, with $0.81 \pm 0.01$ for \texttt{HD 163296} to $0.63 \pm 0.01$ for \texttt{HD 100546} until $0.4 \pm 0.05$ for \texttt{HD 169142} again the lowest fraction as it was seen in the H-band\footnote{For K-band: the extended emission for the three objects are zero for \texttt{HD 163296} and \texttt{HD 100546}. For \texttt{HD 169142} the value is $f_{ext}=0.09$.}. Our models show a fraction of disk K-band emission ranging from $0.46$ for model \texttt{M001} to $0.51$ in model \texttt{M03}. Overall, the fraction of disk emission in the K-band is still higher in the observed disks than in our models, except for the system \texttt{HD 169142}, which shows a value similar to model \texttt{M001}. This is consistent with the median SED showing a higher K-band flux than our models. We summarize: the K-band half-light radius matches our models better, especially for the low magnetized models, which could indicate that the shape, position, and dust temperature could represent the observed systems. Still, an additional hot emission component is needed within the dust rim to explain the H-band half-light radius and the total flux in both H and K bands. 

\subsection{N-band comparison}
We focus on the observations taken with MIDI/VLTI by \citet{men15} for the N-band comparison. At N-band MIDI, the observations recovered the total flux and the half-light radius. We normalized the total flux to a distance of 100~pc as we did for the median SED. One quickly notices a much larger spread in the N-band emission for our selection of the three Herbig-type systems compared to our models. \texttt{HD 100546} has the largest N-band flux, with $69.1 \pm 1.5$~Jy and a half-light radius of $21.3 \pm 1.1$~au. In contrast, \texttt{HD 163296} shows a half-light radius of $0.82 \pm 0.05$~au and a flux of $16.3 \pm 0.6$~Jy. Further \texttt{HD 169142} shows only a flux of $0.92 \pm 0.26$~Jy with a half-light radius of $9.5 \pm 2.3$~au. For all our models, the N-band half-light radius remains inside 1~au, ranging from $0.79$~au for model \texttt{M001} to $0.93$~au for model \texttt{M03}. As previously reported, the magnetic fields affect mostly the mid-infrared flux, including the N-band flux. We obtain a flux of $25.7$~Jy for model \texttt{M001} in the N-band, which increases for higher magnetization to $36.2$~Jy for model \texttt{M03}. Finally, we note again that the median SED showed a total N-band flux of $28 \pm 15$~Jy, which compares very well to our fiducial model. The comparison shows that the system \texttt{HD 163296} best fits our model and the median SED. Given the large deviations for the N-band results for the systems \texttt{HD 100546} and \texttt{HD 169142}, we believe that these systems could be more complex, e.g., either being a circumbinary disk or having a dust-depleted inner disk. \citet{men15} also argue that both systems \texttt{HD 100546} and \texttt{HD 169142} have wide dust gaps affecting their N-band emission. 
\begin{figure}
\resizebox{\hsize}{!}{\includegraphics{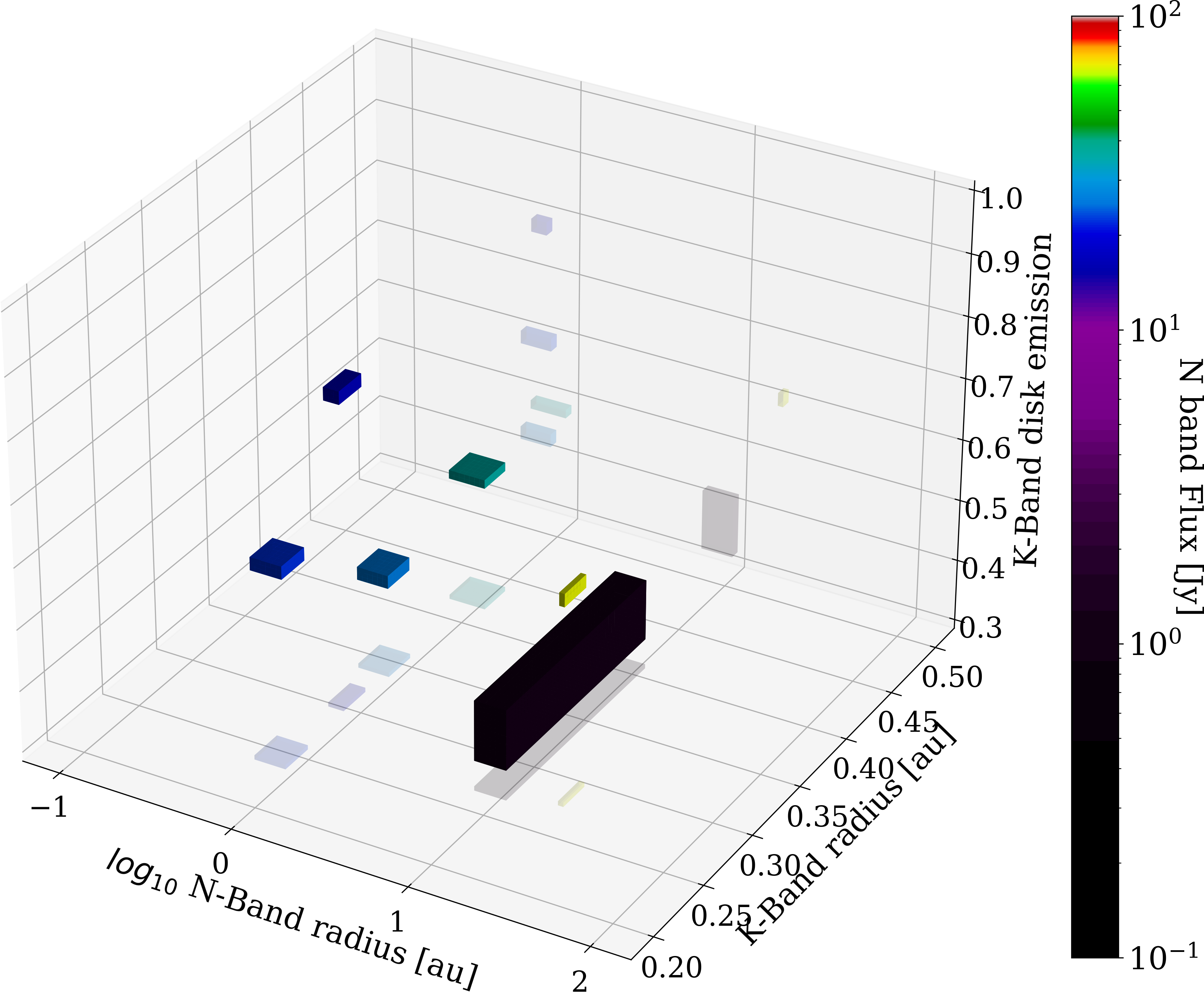}}
\caption{Visualizing the entries in Table~\ref{tab:comp}. For the observed disks, the position and size of the block size represent the corresponding value and error. For the model entries, we set a constant block size (the three equally sized blocks). The color of the block represents the N-band flux normalized at 100 pc for each object and model.}  
\label{fig:comp}
\end{figure}

We conclude the last two subsections by showing Fig.~\ref{fig:comp}, which summarizes the results in the K and N-bands and compares the observations. The system \texttt{HD 169142} (black cuboid) shows the largest error in the K-band half-light radius, still comparing well with the K-band half-light radius of our models. This system shows the lowest N-band emission (color), which might be a combination of having a low accretion rate and the inner disk regions being depleted in dust or shadowed, indicated by the large half-light radius of $\approx 9.5$~au in the N-band. The system \texttt{HD 100546} (yellow cuboid), in contrast, has a relatively large accretion rate and shows the highest N-band flux, together with a colossal N-band half-light radius of $\approx 21.3$~au, indicating a dust-depleted inner disk region. The errors in the H and K-bands are minimal, with the K-band radius also being close to the values from our models. The system \texttt{HD 163296} (dark blue cuboid - top left), with a similar accretion rate as \texttt{HD 100546}, compares best to our models. The half-light radius in the K-band ($\approx 0.3$~au), the N-band half-light radius ($\approx 0.82$~au), and the total N-band flux ($\approx 16$~Jy) are very close to our models, especially the low magnetized model \texttt{M001}.

\begin{figure}
\resizebox{\hsize}{!}{\includegraphics{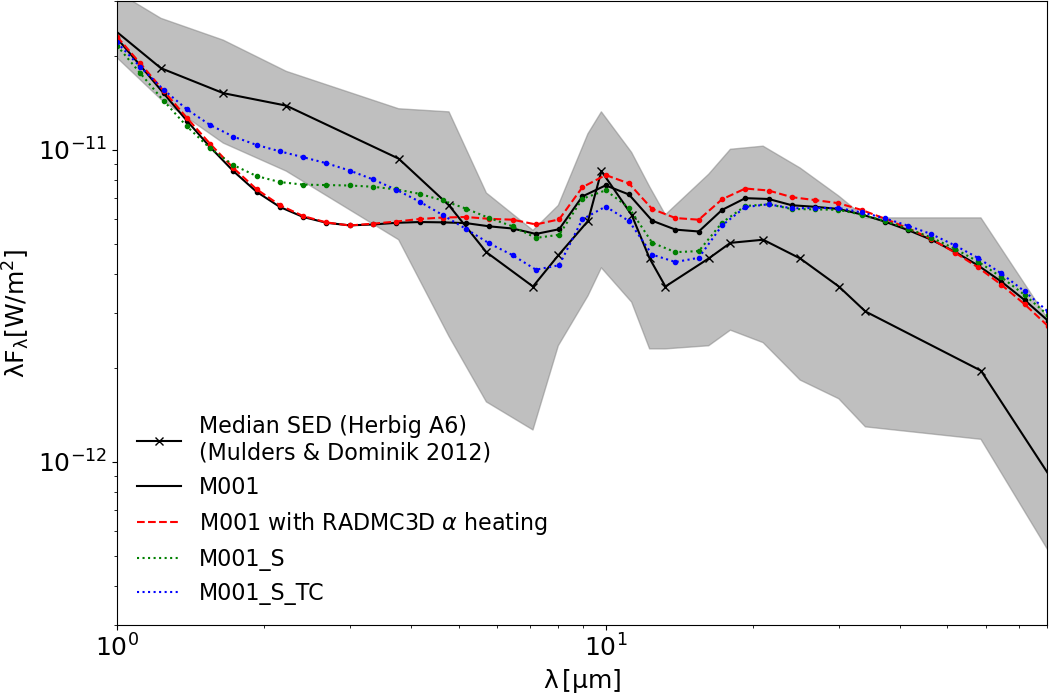}}
\vspace{0.5cm}
\resizebox{\hsize}{!}{\includegraphics{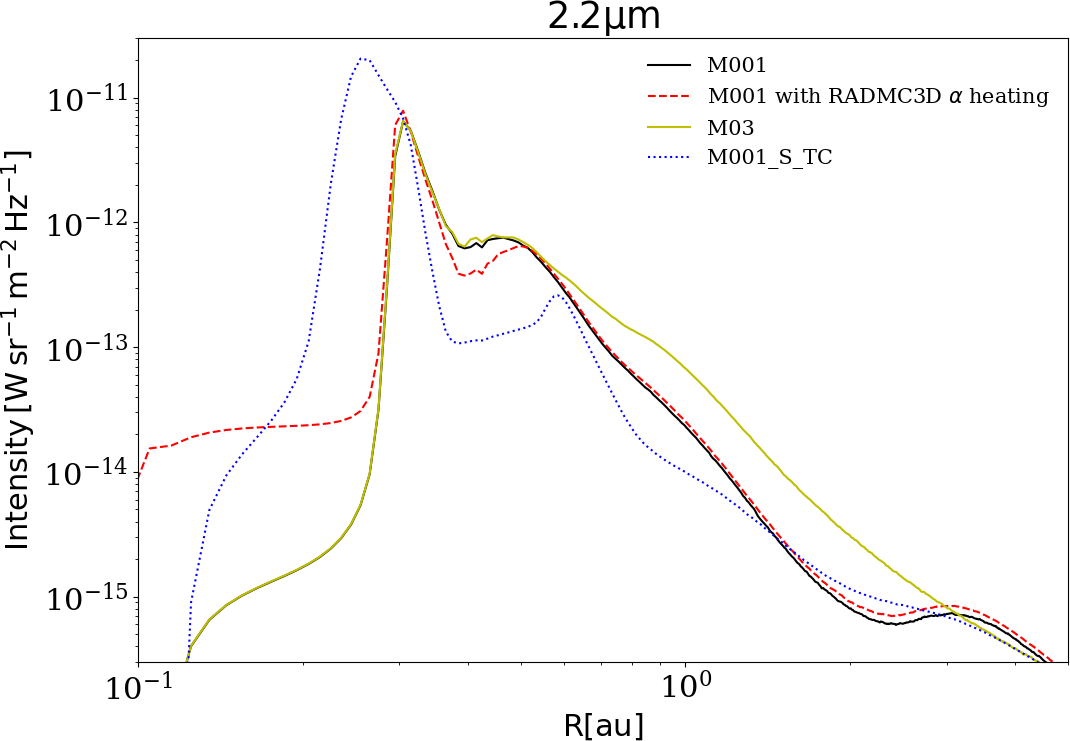}}
\caption{Top: SED for additional models: with accretion heating (red dashed), with only silicate grains (green dotted), and with only silicates with a constant and high ($T_{\rm ev}=1500K$) sublimation temperature \texttt{M001\_S\_TC} (blue dotted). The red dashed line was calculated based on model \texttt{M001}, including accretion heating in the \texttt{RADMC3D} calculation. Bottom: Radial intensity profile { along the major axis (Y=0)} for models \texttt{M001}, \texttt{M03},  \texttt{M001} with accretion heating in \texttt{RADMC3D} and the model \texttt{M001\_S\_TC} {at K-band (2.2~\mum)}.} 
\label{fig:sed_acc}
\end{figure}

\begin{figure*}[t]
\resizebox{\hsize}{!}{\includegraphics{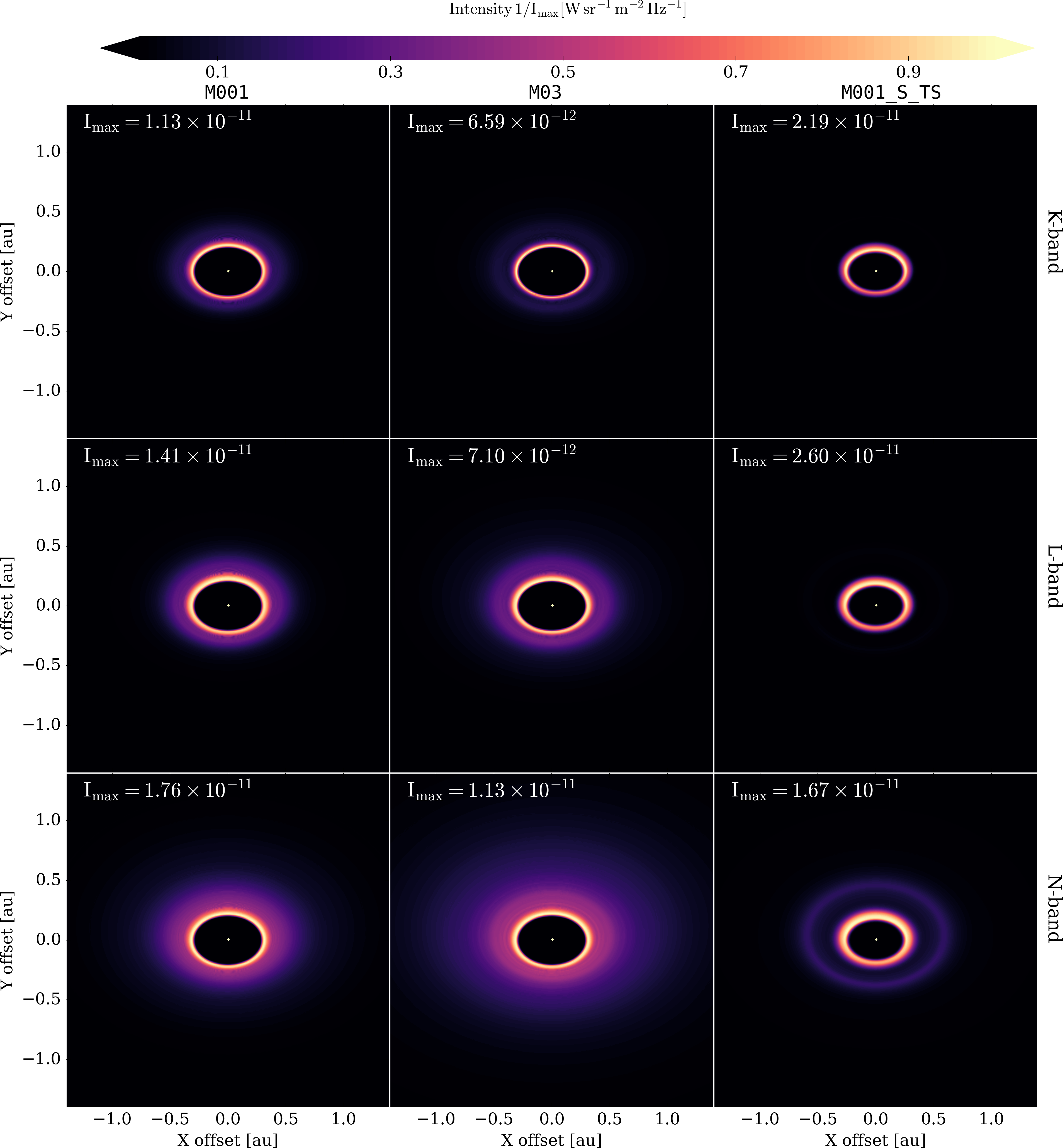}}
\caption{ { 2D intensity map calculated for an inclination of 45$^\circ$ at K-band (2.2~\mum - top), L-band (3.45 \mum - middle) and N-band (10 \mum - bottom) using a linear colormap and normalized over the individual peak intensity in units of $\rm W sr^{-1} m^{-2}Hz^{-1}$.  From left to right: model \texttt{M001}, \texttt{M03} and \texttt{M001\_S\_TS}.}} 
\label{fig:rad_int_2drim}
\end{figure*}

\subsection{H-band emission and effect of accretion heating} \label{sec:intprof}
In our new models, we have shown that adding multiple grain species with realistic sublimation temperatures and abundance makes the rim smoother. However, even though corundum grains survive at higher temperatures, they do not increase the H-band flux. On the contrary, the H-band flux decreases, as the rim becomes broader and more smooth. The effect can be seen in Fig.~\ref{fig:sed_acc}, top. The green dotted line shows that a more compact and less smooth rim helps in increasing the H and K-band flux. Having only silicate grains, the H-band flux increases as the rim becomes less smooth and more wall-like. We added one additional model \texttt{M001\_S\_TC}, similar to \citet{chr24}, to include a single dust component with a constant condensation temperature of 1550~K. This configuration leads to a more compact and hotter dust rim. Such a more wall-like structure was also recommended by \citet{chr24} to fit better the visibility curves for \texttt{HD 163296}. For this model \texttt{M001\_S\_TC}, the SED profile looks very similar to the median SED; compare the blue line in Fig.~\ref{fig:sed_acc}, top. When comparing the half-light radius model \texttt{M001\_S\_TC} can only fit the H-band half-light radius with $a_\mathrm{H}=0.24$, which compares well with the systems \texttt{HD 163296} and \texttt{HD 100546}. However, the half-light radius of the K-band remains too small compared to observations. We, therefore, conclude that a better fitting model would be similar to \texttt{M001} with additional H-band flux of unknown origin without affecting the overall smooth rim structure much. In the discussion section, we present additional RADMC3D models that could produce this additional flux. The details of these pure silicate models can be found in Appendix \texttt{C}. In the previous subsections, we ignored the accretion heating in the RADMC3D models. To check the influence of the viscous heating onto the SED, we incorporate viscous heating into \texttt{RADMC3D} by using the local $\alpha_{heat}$ to calculate the internal heating using the {\it heatsource.inp} file. Here, we implement a fixed $\alpha$ heating profile depending on the radius using 
\begin{equation} \alpha_\mathrm{heat}=\frac{1}{2} ( \alpha_\mathrm{MRI} - \alpha_\mathrm{DZ}) \left [ 1.0 - \tanh \left ( \frac{r-0.3\, \mathrm{au}}{0.1\, \mathrm{au}} \right  ) \right ] + \alpha_\mathrm{DZ}.\end{equation} to also capture the stronger heating inwards the dust rim expected by the MRI. The value of $0.3$~au was chosen to represent the MRI transition of the reference model. The profile is static to prevent any dynamic feedback. The transition range was chosen here over $0.1$~au. A model SED with such a viscous heating is shown in Fig.~\ref{fig:sed_acc}; see the red dashed line. There is nearly no difference compared to the passive model, with only a slight increase in the flux between 5 and 20 \mum. E.g., we see an 8\%  flux increase with accretion heating at 10 \mum. This can be explained by accretion heating, which increases only the midplane temperature. It does not strongly affect the temperature layer directly below the irradiation absorption, which contributes most to the NIR emission. Further, the MRI heated, dust-free, optical thin regime contributes only a little to the total flux, see Fig.~\ref{fig:sed_acc}, bottom.

The radial intensity profile at the K-band is shown in Fig.~\ref{fig:sed_acc}, bottom. The intensity peak is emitted from the 'nose' of the dust rim, located at the midplane and around 0.35~au for most models. The accretion heating increases the emission inside the rim, see Fig.~\ref{fig:sed_acc}, bottom, but not strong enough to change the total flux, see Fig.~\ref{fig:sed_acc}, top. The profile remains very similar to the passively heated model. The radial intensity profile of the higher magnetized model \texttt{M03} shows that the flux increase comes from the region radially outward 0.6~au, which can be explained by the slightly puffed-up surface for irradiation absorption. For the model \texttt{M001\_S\_TC} the emission is more compact and overall higher due to the higher temperature of the 'wall-like' rim.

\subsection{Synthetic images at K, L and N-band}
The 2D images, calculated with an inclination of 45 degrees at K-band, L-band and N-band are summarized in Fig.~\ref{fig:rad_int_2drim}. { The low and moderate magnetization models \texttt{M001} and \texttt{M03} show very similar emission profiles in the K and L-band.} They show the peak emission from a narrow ring with a broader, less intense outer ring. { In the N-band, model \texttt{M03} shows a more extended emission area due to the small changes in the flaring angle.} 
For comparison, we also show the model \texttt{M001\_S\_TS} with one dust component and a fine-tuned dust sublimation temperature independent of the local pressure. This model image shows a more compact, broader, and brighter ring of emission. In the Appendix \texttt{E}, we show further an image for the very magnetized model \texttt{M3}.

\subsection{Toy model accounting for the emission deficit by adding hot dust emission}
\begin{figure}
\resizebox{\hsize}{!}{\includegraphics{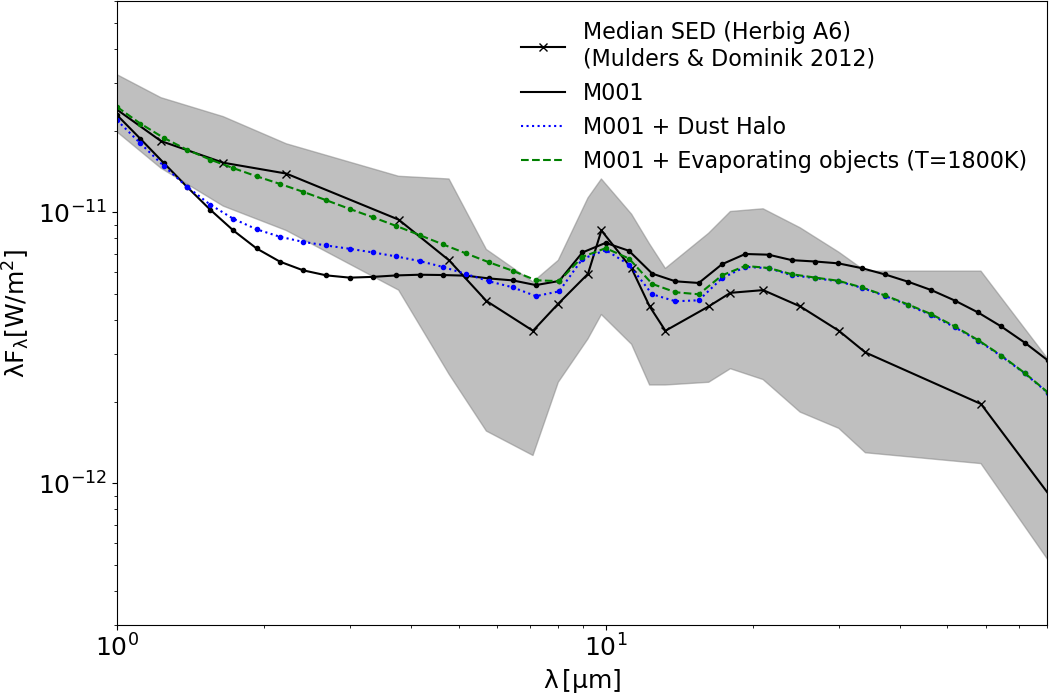}}
\caption{Similar to Fig.~\ref{fig:sed} showing the SED for two additional models, including a dust halo (blue dotted) and evaporating objects (green dashed line). The black solid line (crosses) and grey shaded color show the median SED and one-sigma error constructed by \citet{mul12}. All values shown here were normalized to a distance at 100~pc.} 
\label{fig:sed_halo}
\end{figure}

In this section, we present two additional \texttt{RADMC3D} models. The goal here is to boost the H-band emission without changing the physical structure of our models. In the first model,
we add a small amount of corundum dust to mimic more dust material inside the rim. We added a flat value of the dust to gas mass ratio to the corundum dust density times the gas density $10^{-8} \rho$. This should mimic an optically thin dust halo surrounding the disk. For this dust halo, we neglect dust sublimation for these extra grains. The SED profile of this 'Dust Halo' RADMC3D model is shown in Fig~\ref{fig:sed_halo}, blue dotted line. The H and K-band emission increases due to the higher emission of hot material; however, it still cannot explain the observed values.

For the second model, we assume emission from compact evaporating planetesimals with a surface temperature of 1800~K, a circular emitting area with a radius of 0.0733~au. The choice of the temperature is motivated by the highest refractory material sublimation temperature. The choice of the area is here arbitrary and set to match the emission at 2 $\rm \mu m$. As we would expect large tails of hot dust emission due to the planetesimals' evaporation, it is not easy to estimate the amount and mass of planetesimal material that could cause this emission. We expect the evaporation tail, and therefore the emitting area, to be several orders of magnitude larger than the planetesimal's area itself \citep{bie63}. For this 'Evaporating objects' model, we add a point source of stellar emission in the {\it stars.inp} file of the \texttt{RADMC3D} setup. The exact position is at the center in the X-Y plane and in the Z-direction 0.015~au above the central star, which is a constraint for the current axisymmetric setup and version of \texttt{RADMC3D}. We show and compare the SED profile in Fig~\ref{fig:sed_halo}, green dashed line. The 'Evaporating objects' model, which assumed an additional continuum source to mimic evaporating planetesimals, can match the H and K band emission flux well.  
We note that both models require fine-tuning. The extra dust is fine-tuned as it cannot be too large, as it otherwise becomes optically thick for the irradiation, which reduces the temperature in the inner regions. For the small objects model, we needed to fine-tune the area to match the emission. 

\section{Discussion}
\subsection{How broad is the $\alpha$ transition at the inner dead-zone edge.}
One important property of the inner dead-zone edge is the steepness of the surface density rise. A steeper rise in surface density leads to larger outward migration torque onto planets trapped at this location \citep{mas06,flo19,chr22}. At the same time, the steepness of the surface density rise determines the strength of the radial pressure gradient and so the efficiency of concentrating pebbles \citep{izi21}. This steepness is determined by the drop of the MRI activity, which depends on the strengths of the non-ideal diffusivity of Ohmic, Ambipolar, and Hall. 
As a guideline, \citet{des15} (see Fig. 17 for the midplane) showed that the Ohmic and Ambipolar Elsasser number increases from 0.1 to 1.0 between 900~K and 950~K, pointing to a transition of around 50~K. However, this depends on many parameters, like the strength of the magnetic field, the local ionization rate by radiation, the local gas and dust density, and the dust size \citep{del22}. New models of ionization focusing on the inner disk \citet{wil25} emphasize the importance of the grain size distribution, also showing a very steep drop in resistivity with temperature, however, not specifying the Elsasser number. Global 3D radiation non-ideal MHD simulations also point to a very sharp drop in MRI activity \citep{flo17} between 900~K and 1000~K, with an $\alpha$ drop by at least 2 orders of magnitude. We summarize that most of the works point to a very sharp ionization transition between 100~K and below, close to the resolution limit of current 2D or 3D radiation MHD models.

\subsection{Constraining the magnetic flux in the inner disk} 
With our models, one goal was to answer the question about the strength of the magnetic field in the inner disk based on the current observational constraints. { The goal is based on the idea that the disk becomes more puffed-up due to magnetic pressure support, and so the thermal infrared emission substantially increases.} Our results show that this effect occurs at relatively strong magnetic field strengths for the models \texttt{M03} and beyond, corresponding to plasma beta values of around $\approx 10^3$. Such a substantial excess in emission is not seen in the median SED. We therefore argue that most of the observed young Herbig disks are weakly magnetized with beta values $\gtrapprox 10^4$.
We remember the limitation of our models, especially for the strongly magnetized models in Appendix \texttt{E}. Dynamical simulations will here show a substantial increase in the accretion rate \citep{les23}, reducing the density and changing the temperature structure. Future models should include the magnetic torque and the corresponding accretion stress. 

\subsection{Location of the magnetic wind base}
The wind base of the magnetic-driven wind can be approximated in our models with the $\tau_\mathrm{r}=1$ layer as most of the FUV radiation is absorbed { close to this layer}; see blue line in Fig.~\ref{fig:tempmag}. By comparing model \texttt{M001} to model \texttt{M03}, we see that the magnetization at this location changes from a weakly magnetized case ($\beta > 1 $) in model \texttt{M001} to a stronger magnetized case ($\beta < 1$) in model \texttt{M03} at this height. However, even with this change, the location in height for both $\tau_\mathrm{r}=1$ and $\beta=1$ remains close by, which explains that the resulting emission profiles are very similar. Our results show that as long as the FUV or irradiation absorption layers remain below or close to the magnetized layer ($\beta=1$ region), we would not expect a substantial change in the NIR emission. However, we notice a stronger effect in the N-band emission, which is due to the change in the grazing angle. The increase in the N-band flux for higher magnetization was seen both in the images and in the total flux, see Table~\ref{tab:comp} and Fig.~\ref{fig:rad_int_2drim}. 

\subsection{The nature of H-band emission and opacity sources inside the silicate sublimation.}
Our results show that including magnetization and multi-dust species affects the density and temperature structure. Although the infrared flux increases for stronger magnetic fields, the results show that magnetic fields alone cannot explain the excess in the H-band flux. \citet{tur14} were able to find a better match to the observed H-band flux using a magnetically supported disk. However, they set the inner dust rim location by hand without including dust sublimation. Further, they used a density fit to model the effect of the magnetic field, and they modeled only one field strength. Their models also were overshooting the emission flux at around 6 to 7 \mum, similar to our models with a field strength above $\rm B_z = 0.3\, G$ at 1~au (see Appendix \texttt{E}).

Adding corundum grains makes the rim smoother and radially more extended, which has the opposite effect of a sharp wall and even reduces the flux in the H-band (see Fig.~\ref{fig:sed_acc}). Another potential opacity source inside the silicate sublimation could come from graphite. In this work, we do not consider graphite grains, as several studies have pointed out that it is not the main carbon carrier in protoplanetary disks. Most carbon is supposed to be locked in amorphous carbon or PAH \citep{dra84,ser08,and17}. Furthermore, graphite is not included in the reference model of \texttt{GGChem}, which we choose for our model, see \citet{woi18}. They argue that most of the carbon is being lost at around 500~K at the soot line, see also \citet{ber23}. A recent work by \citet{and17} mentioned that refractory carbon might be destroyed before reaching the inner disk. However, we note that most of the works still focus on the main carbon carriers in amorphous grains and PAHs. Adding a small amount of graphite could lead to an optically thin heated layer, which would be an interesting study for the future.

Another source of opacity could come from various metal lines in the gas phase \citep{dul10}. By integrating over these lines with high spectral resolution \citet{mal14} showed that metal lines in the gas phase can contribute significantly to the mean Planck opacity. Such high opacity would block most of the stellar irradiation and shift the dust rim radially inwards. Realistic photochemistry \citep{kam17} and self-shielding rates might help to overcome this. In contrast, the Rosseland mean opacities are expected to remain small for the gas opacity, with values similar to those taken for this work (see more in Appendix B1 by \citet{flo19}).

By fine-tuning and modifying the sublimation temperature and using a single dust component, we can almost match the H-band flux of the median SED and the H-band half-light radius (see also \citet{chr24}). However, this tuned model is inconsistent with the observed K and L-band half-light radius, and it ignores the sublimation temperature derived by \texttt{GGchem}. To match the H-band flux without affecting the rest of the disk structure of our reference model, we need an additional continuum component, which is geometrically thin and so does not block the irradiation onto the dust rim. Larger objects, e.g., a swarm of km-sized evaporating objects on eccentric orbits, might be able to contribute an additional hot component.

\section{Conclusion}
This work presents a new class of 2D radiative magneto-hydrostatic multi-dust species disk models for protoplanetary disks. Our reference model follows a Herbig-type star-disk system with $L_* = 21 {L_\sun}$ and mass accretion rate of $ \dot{M}=1.0 \times 10^{-8} M_\sun/{\rm yr}$. Dust sublimation is determined using \texttt{GGchem} using solar abundance ratios for corundum, iron, forsterite, and enstatite grains at different temperatures and densities. Opacities are determined using the \texttt{Optool} package for each dust species. Frequency-dependent irradiation, viscous heating, and temperature-dependent Planck and Rosseland opacities are implemented. 
By changing the strength of the magnetic field, we obtain several temperature and density profiles with different magnetization levels. Post-processing with Monte-Carlo radiative transfer, we determine the SED, the half-light radius in the H, K, and N-bands, as well as the disk emission fraction for these wavelengths. We present synthetic images and also compare our models to the results of detailed H, K, and N-band interferometric observations of the systems \texttt{HD 163296}, \texttt{HD 169142} and \texttt{HD 100546} based on the previous surveys by the PIONIER and GRAVITY instruments at the VLTI. Our key results are the following:

\begin{itemize}

    \item Adding multiple dust components with different sublimation temperatures makes the inner dust rim broader and smoother. Corundum grains survive closer inside the inner rim, contributing as hot and optically thin dust emission in front of the dust rim. The optical thick rim starts at around 0.35~au for our star and disk parameter choice. In our models, the magnetic fields mainly affect the upper layers in the disk where the magnetic pressure dominates, increasing the surface where stellar irradiation is absorbed. 

    \item For our typical Herbig model, we observe a pressure bump at 0.62~au and so an efficient trap for inward drifting pebbles. The effect of including accretion heating and a temperature-dependent dust opacity leads to variations in the steepness of the temperature drop along the radius. These variations lead to perturbations in the radial profile of H/R and result in several shadowed regions.
    
    \item Based on the detailed comparison with observations in the H, K, and N-band, we find that magnetic field strengths of 0.1 Gauss at 1au and below fit best to the median SED. These magnetic field strengths correspond to midplane plasma beta values of around $2 \times 10^4$ and higher. This magnetic field strength is high enough to enable the model's mass accretion rate by magnetic-driven winds. 
    
    \item The effect of the magnetic fields gets important once the plasma beta unity layer is vertically less extended than the irradiation layer (the height of the $\beta=1$ line compared to the $\tau_\mathrm{r} =1$ line). In contrast, the magnetic field strength of 0.3 Gauss and above at 1au shows an excess emission in the mid-infrared.

%
%
%\end{itemize}
%We compared, in more detail, total N-Band flux and half-light radii for H, K, and N Band of 3 Herbig stars with similar stellar luminosities as our models, namely \texttt{HD 163296}, \texttt{HD 169142} and \texttt{HD 100546}. All flux results were normalized to a distance of 100~pc.

%\begin{itemize}
    \item The observed H-band flux is still higher than in our models. Additional dust species make the rim smoother and so reduce the H-band emission compared to a single dust model with a steeper rim shape. Our results show that the magnetic pressure-supported atmosphere increases the emission in the mid-infrared (3-9 $\rm \mu m$) but less in the near-infrared ($\rm \le 2~\mu m$), as most of the H-band flux is coming from a region near the midplane, which remains unaffected by magnetic fields. 
    
    \item The observed half-light radius of the H-band emission is closer to the star and inwards the dust rim. The observations show a half-light radius of $a_\mathrm{H} \leq 0.26$~au while our models show $a_\mathrm{H} \geq 0.31$. The results indicate that a geometrically thin, hot continuum emission is still missing. We are able to increase the H-band emission using an additional optical thin, non-sublimating dust component. We are able to match the H-band and K-band emission of the median SED by fine-tuning the area of an additional compact emission at $T=1800K$, which should mimic a swarm of eccentric, larger evaporating objects inside the rim location.
    
    \item The K-band half-light radius best fits our disk models, especially for low and moderate magnetization levels. Higher magnetization makes the K-band emission broader and shifts the half-light radius outwards. The total K-band flux from observations is also higher than our models, as in the case of the H-band. 

    \item  The N-band flux of $\rm 25.7~Jy$ for our low magnetized model matches best the median SED value. In contrast, we see a substantial spread in the N-band flux for the three specific disk systems, ranging over a factor of 75, with \texttt{HD 169142} having the lowest value with $\rm 0.92 \pm 0.26~Jy$ and \texttt{HD 100546} the highest with $\rm 69.1 \pm 1.5~Jy$. The system \texttt{HD 163296} best fits the low and moderate magnetization models when comparing K-band radius, N-band radius, and total N-band flux. { With increasing magnetization, the N-band flux increases in our models, which might help to constrain disk magnetization in the future.}

    \item We find that the effect of viscous heating on the near and mid-infrared emission remains low. The viscous heating mainly affects the midplane temperatures. While the water snow line is pushed outward, the temperature in the upper disk layers remains primarily unchanged. A viscously heated, Monte-Carlo radiative transfer disk model shows an increase of N-band flux by only 8\% compared to the passively heated disk model. 
    
\end{itemize}

Finally, we find that very high magnetization levels change the disk's structure to an envelope. In this case, the vertical density profile becomes flat due to the dominant magnetic pressure, and the temperature falls off with the radial distance to the star. Future work should perform such model cases in dynamical simulations to investigate the structure, which might be relevant for the very early phase of star and disk formation. 

\begin{acknowledgements}
The authors thank Karine Perraut, Catherine Dougados, and Lucas Labadie for their discussions and feedback on interpreting the VLTI results during the YSO Gravity meeting in 2024. We also thank Jean-Philip Berger and John Monnier for their discussions and for interpreting the CHARA results.  This research was supported in part by the grant NSF PHY-2309135 to the Kavli Institute for Theoretical Physics (KITP). M. F. has received funding from the European Research Council (ERC) under the European Union's Horizon 2020 research and innovation program (grant agreement No. 757957). The work of O.C. was supported by the Czech Science Foundation (grant 21-23067M) and the Charles University Research Centre program (No. UNCE/24/SCI/005).  T.U. acknowledges the support of the DFG-Grant, project number 465962023. J. V. is funded by the Hungarian NKFIH OTKA project no. K-132406 and no. K-147380. J. V. acknowledges support from the Fizeau Exchange Visitors program. The research leading to these results has received funding from the European Union’s Horizon 2020 research and innovation program under Grant Agreement 101004719 (ORP). This project has received funding from the European Research Council (ERC) under the European Union’s Horizon 2020 research and innovation program (PROTOPLANETS, grant agreement No. 101002188). This work was also supported by the NKFIH NKKP grant ADVANCED 149943 and the NKFIH excellence grant TKP2021-NKTA-64. Project no.149943 has been implemented with the support provided by the Ministry of Culture and Innovation of Hungary from the National Research, Development and Innovation Fund, financed under the NKKP ADVANCED funding scheme.
\end{acknowledgements}

\bibliographystyle{aa}
\bibliography{MultiDRim}

\begin{thebibliography}{80}
\expandafter\ifx\csname natexlab\endcsname\relax\def\natexlab#1{#1}\fi

\bibitem[{{Anderson} {et~al.}(2017){Anderson}, {Bergin}, {Blake}, {Ciesla},
  {Visser}, \& {Lee}}]{and17}
{Anderson}, D.~E., {Bergin}, E.~A., {Blake}, G.~A., {et~al.} 2017, \apj, 845,
  13

\bibitem[{{Bai} \& {Goodman}(2009)}]{bai09}
{Bai}, X.-N. \& {Goodman}, J. 2009, \apj, 701, 737

\bibitem[{{Bans} \& {K{\"o}nigl}(2012)}]{ban12}
{Bans}, A. \& {K{\"o}nigl}, A. 2012, \apj, 758, 100

\bibitem[{{Batalha} {et~al.}(2013){Batalha}, {Rowe}, {Bryson}, {Barclay},
  {Burke}, {Caldwell}, {Christiansen}, {Mullally}, {Thompson}, {Brown},
  {Dupree}, {Fabrycky}, {Ford}, {Fortney}, {Gilliland}, {Isaacson}, {Latham},
  {Marcy}, {Quinn}, {Ragozzine}, {Shporer}, {Borucki}, {Ciardi}, {Gautier},
  {Haas}, {Jenkins}, {Koch}, {Lissauer}, {Rapin}, {Basri}, {Boss}, {Buchhave},
  {Carter}, {Charbonneau}, {Christensen-Dalsgaard}, {Clarke}, {Cochran},
  {Demory}, {Desert}, {Devore}, {Doyle}, {Esquerdo}, {Everett}, {Fressin},
  {Geary}, {Girouard}, {Gould}, {Hall}, {Holman}, {Howard}, {Howell},
  {Ibrahim}, {Kinemuchi}, {Kjeldsen}, {Klaus}, {Li}, {Lucas}, {Meibom},
  {Morris}, {Pr{\v s}a}, {Quintana}, {Sanderfer}, {Sasselov}, {Seader},
  {Smith}, {Steffen}, {Still}, {Stumpe}, {Tarter}, {Tenenbaum}, {Torres},
  {Twicken}, {Uddin}, {Van Cleve}, {Walkowicz}, \& {Welsh}}]{bat13}
{Batalha}, N.~M., {Rowe}, J.~F., {Bryson}, S.~T., {et~al.} 2013, \apjs, 204, 24

\bibitem[{{Bergin} {et~al.}(2023){Bergin}, {Kempton}, {Hirschmann},
  {Bastelberger}, {Teal}, {Blake}, {Ciesla}, \& {Li}}]{ber23}
{Bergin}, E.~A., {Kempton}, E. M.~R., {Hirschmann}, M., {et~al.} 2023, \apjl,
  949, L17

\bibitem[{{Biermann}(1963)}]{bie63}
{Biermann}, L. 1963, \ssr, 1, 553

\bibitem[{{Bitsch} {et~al.}(2013){Bitsch}, {Crida}, {Morbidelli}, {Kley}, \&
  {Dobbs-Dixon}}]{bit13a}
{Bitsch}, B., {Crida}, A., {Morbidelli}, A., {Kley}, W., \& {Dobbs-Dixon}, I.
  2013, \aap, 549, A124

\bibitem[{{Boley} {et~al.}(2014){Boley}, {Morris}, \& {Ford}}]{bol14}
{Boley}, A.~C., {Morris}, M.~A., \& {Ford}, E.~B. 2014, \apjl, 792, L27

\bibitem[{{Cecil} \& {Flock}(2024)}]{cec24}
{Cecil}, M. \& {Flock}, M. 2024, \aap, 692, A171

\bibitem[{{Chatterjee} \& {Tan}(2014)}]{cha14}
{Chatterjee}, S. \& {Tan}, J.~C. 2014, \apj, 780, 53

\bibitem[{{Chrenko} {et~al.}(2022){Chrenko}, {Chametla}, {Nesvorn{\'y}}, \&
  {Flock}}]{chr22}
{Chrenko}, O., {Chametla}, R.~O., {Nesvorn{\'y}}, D., \& {Flock}, M. 2022,
  \aap, 666, A63

\bibitem[{{Chrenko} {et~al.}(2024){Chrenko}, {Flock}, {Ueda}, {M{\'e}rand},
  {Benisty}, \& {Chametla}}]{chr24}
{Chrenko}, O., {Flock}, M., {Ueda}, T., {et~al.} 2024, \aj, 167, 124

\bibitem[{{Colavita} {et~al.}(2013){Colavita}, {Wizinowich}, {Akeson},
  {Ragland}, {Woillez}, {Millan-Gabet}, {Serabyn}, {Abajian}, {Acton},
  {Appleby}, {Beletic}, {Beichman}, {Bell}, {Berkey}, {Berlin}, {Boden},
  {Booth}, {Boutell}, {Chaffee}, {Chan}, {Chin}, {Chock}, {Cohen}, {Cooper},
  {Crawford}, {Creech-Eakman}, {Dahl}, {Eychaner}, {Fanson}, {Felizardo},
  {Garcia-Gathright}, {Gathright}, {Hardy}, {Henderson}, {Herstein}, {Hess},
  {Hovland}, {Hrynevych}, {Johansson}, {Johnson}, {Kelley}, {Kendrick},
  {Koresko}, {Kurpis}, {Le Mignant}, {Lewis}, {Ligon}, {Lupton}, {McBride},
  {Medeiros}, {Mennesson}, {Moore}, {Morrison}, {Nance}, {Neyman}, {Niessner},
  {Paine}, {Palmer}, {Panteleeva}, {Papin}, {Parvin}, {Reder}, {Rudeen},
  {Saloga}, {Sargent}, {Shao}, {Smith}, {Smythe}, {Stomski}, {Summers},
  {Swain}, {Swanson}, {Thompson}, {Tsubota}, {Tumminello}, {Tyau}, {van Belle},
  {Vasisht}, {Vause}, {Vescelus}, {Walker}, {Wallace}, {Wehmeier}, \&
  {Wetherell}}]{col13}
{Colavita}, M.~M., {Wizinowich}, P.~L., {Akeson}, R.~L., {et~al.} 2013, \pasp,
  125, 1226

\bibitem[{{D'Angelo} \& {Bodenheimer}(2013)}]{dan13}
{D'Angelo}, G. \& {Bodenheimer}, P. 2013, \apj, 778, 77

\bibitem[{{Delage} {et~al.}(2022){Delage}, {Okuzumi}, {Flock}, {Pinilla}, \&
  {Dzyurkevich}}]{del22}
{Delage}, T.~N., {Okuzumi}, S., {Flock}, M., {Pinilla}, P., \& {Dzyurkevich},
  N. 2022, \aap, 658, A97

\bibitem[{{Desch} \& {Turner}(2015)}]{des15}
{Desch}, S.~J. \& {Turner}, N.~J. 2015, \apj, 811, 156

\bibitem[{{Draine} \& {Lee}(1984)}]{dra84}
{Draine}, B.~T. \& {Lee}, H.~M. 1984, \apj, 285, 89

\bibitem[{{Dr{\k{a}}{\.z}kowska} {et~al.}(2023){Dr{\k{a}}{\.z}kowska},
  {Bitsch}, {Lambrechts}, {Mulders}, {Harsono}, {Vazan}, {Liu}, {Ormel},
  {Kretke}, \& {Morbidelli}}]{dra23}
{Dr{\k{a}}{\.z}kowska}, J., {Bitsch}, B., {Lambrechts}, M., {et~al.} 2023, in
  Astronomical Society of the Pacific Conference Series, Vol. 534, Protostars
  and Planets VII, ed. S.~{Inutsuka}, Y.~{Aikawa}, T.~{Muto}, K.~{Tomida}, \&
  M.~{Tamura}, 717

\bibitem[{{Dullemond}(2012)}]{dul12}
{Dullemond}, C.~P. 2012, {RADMC-3D: A multi-purpose radiative transfer tool},
  astrophysics Source Code Library

\bibitem[{{Dullemond} {et~al.}(2001){Dullemond}, {Dominik}, \& {Natta}}]{dul01}
{Dullemond}, C.~P., {Dominik}, C., \& {Natta}, A. 2001, \apj, 560, 957

\bibitem[{{Dullemond} \& {Monnier}(2010)}]{dul10}
{Dullemond}, C.~P. \& {Monnier}, J.~D. 2010, \araa, 48, 205

\bibitem[{{Dzyurkevich} {et~al.}(2010){Dzyurkevich}, {Flock}, {Turner},
  {Klahr}, \& {Henning}}]{dzy10}
{Dzyurkevich}, N., {Flock}, M., {Turner}, N.~J., {Klahr}, H., \& {Henning}, T.
  2010, \aap, 515, A70

\bibitem[{{Fabrycky} {et~al.}(2014){Fabrycky}, {Lissauer}, {Ragozzine}, {Rowe},
  {Steffen}, {Agol}, {Barclay}, {Batalha}, {Borucki}, {Ciardi}, {Ford},
  {Gautier}, {Geary}, {Holman}, {Jenkins}, {Li}, {Morehead}, {Morris},
  {Shporer}, {Smith}, {Still}, \& {Van Cleve}}]{fab14}
{Fabrycky}, D.~C., {Lissauer}, J.~J., {Ragozzine}, D., {et~al.} 2014, \apj,
  790, 146

\bibitem[{{Flock} {et~al.}(2013){Flock}, {Fromang}, {Gonz{\'a}lez}, \&
  {Commer{\c c}on}}]{flo13}
{Flock}, M., {Fromang}, S., {Gonz{\'a}lez}, M., \& {Commer{\c c}on}, B. 2013,
  \aap, 560, A43

\bibitem[{{Flock} {et~al.}(2016){Flock}, {Fromang}, {Turner}, \&
  {Benisty}}]{flo16}
{Flock}, M., {Fromang}, S., {Turner}, N.~J., \& {Benisty}, M. 2016, \apj, 827,
  144

\bibitem[{{Flock} {et~al.}(2017){Flock}, {Fromang}, {Turner}, \&
  {Benisty}}]{flo17}
{Flock}, M., {Fromang}, S., {Turner}, N.~J., \& {Benisty}, M. 2017, \apj, 835,
  230

\bibitem[{{Flock} {et~al.}(2019){Flock}, {Turner}, {Mulders}, {Hasegawa},
  {Nelson}, \& {Bitsch}}]{flo19}
{Flock}, M., {Turner}, N.~J., {Mulders}, G.~D., {et~al.} 2019, \aap, 630, A147

\bibitem[{{Furlan} {et~al.}(2006){Furlan}, {Hartmann}, {Calvet}, {D'Alessio},
  {Franco-Hern{\'a}ndez}, {Forrest}, {Watson}, {Uchida}, {Sargent}, {Green},
  {Keller}, \& {Herter}}]{Fur06}
{Furlan}, E., {Hartmann}, L., {Calvet}, N., {et~al.} 2006, \apjs, 165, 568

\bibitem[{{Grady} {et~al.}(2007){Grady}, {Schneider}, {Hamaguchi}, {Sitko},
  {Carpenter}, {Hines}, {Collins}, {Williger}, {Woodgate}, {Henning},
  {M{\'e}nard}, {Wilner}, {Petre}, {Palunas}, {Quirrenbach}, {Nuth},
  {Silverstone}, \& {Kim}}]{gra07}
{Grady}, C.~A., {Schneider}, G., {Hamaguchi}, K., {et~al.} 2007, \apj, 665,
  1391

\bibitem[{{GRAVITY Collaboration} {et~al.}(2019){GRAVITY Collaboration},
  {Perraut}, {Labadie}, {Lazareff}, {Klarmann}, {Segura-Cox}, {Benisty},
  {Bouvier}, {Brandner}, {Caratti O Garatti}, {Caselli}, {Dougados}, {Garcia},
  {Garcia-Lopez}, {Kendrew}, {Koutoulaki}, {Kervella}, {Lin}, {Pineda},
  {Sanchez-Bermudez}, {van Dishoeck}, {Abuter}, {Amorim}, {Berger}, {Bonnet},
  {Buron}, {Cantalloube}, {Cl{\'e}net}, {Coud{\'e} Du Foresto}, {Dexter}, {de
  Zeeuw}, {Duvert}, {Eckart}, {Eisenhauer}, {Eupen}, {Gao}, {Gendron},
  {Genzel}, {Gillessen}, {Gordo}, {Grellmann}, {Haubois}, {Haussmann},
  {Henning}, {Hippler}, {Horrobin}, {Hubert}, {Jocou}, {Lacour}, {Le Bouquin},
  {L{\'e}na}, {M{\'e}rand}, {Ott}, {Paumard}, {Perrin}, {Pfuhl}, {Rabien},
  {Ray}, {Rau}, {Rousset}, {Scheithauer}, {Straub}, {Straubmeier}, {Sturm},
  {Vincent}, {Waisberg}, {Wank}, {Widmann}, {Wieprecht}, {Wiest}, {Wiezorrek},
  {Woillez}, \& {Yazici}}]{gra19}
{GRAVITY Collaboration}, {Perraut}, K., {Labadie}, L., {et~al.} 2019, \aap,
  632, A53

\bibitem[{{Hillenbrand} {et~al.}(1992){Hillenbrand}, {Strom}, {Vrba}, \&
  {Keene}}]{hil92}
{Hillenbrand}, L.~A., {Strom}, S.~E., {Vrba}, F.~J., \& {Keene}, J. 1992, \apj,
  397, 613

\bibitem[{{Ida} \& {Lin}(2010)}]{ida10}
{Ida}, S. \& {Lin}, D.~N.~C. 2010, \apj, 719, 810

\bibitem[{{Izidoro} {et~al.}(2021){Izidoro}, {Bitsch}, \& {Dasgupta}}]{izi21}
{Izidoro}, A., {Bitsch}, B., \& {Dasgupta}, R. 2021, \apj, 915, 62

\bibitem[{{Izidoro} {et~al.}(2017){Izidoro}, {Ogihara}, {Raymond},
  {Morbidelli}, {Pierens}, {Bitsch}, {Cossou}, \& {Hersant}}]{izi17}
{Izidoro}, A., {Ogihara}, M., {Raymond}, S.~N., {et~al.} 2017, \mnras, 470,
  1750

\bibitem[{{Kama} {et~al.}(2009){Kama}, {Min}, \& {Dominik}}]{kam09}
{Kama}, M., {Min}, M., \& {Dominik}, C. 2009, \aap, 506, 1199

\bibitem[{{Kamp} {et~al.}(2017){Kamp}, {Thi}, {Woitke}, {Rab}, {Bouma}, \&
  {M{\'e}nard}}]{kam17}
{Kamp}, I., {Thi}, W.~F., {Woitke}, P., {et~al.} 2017, \aap, 607, A41

\bibitem[{{Kutra} {et~al.}(2024){Kutra}, {Wu}, \& {Lithwick}}]{kut24}
{Kutra}, T., {Wu}, Y., \& {Lithwick}, Y. 2024, \apj, 964, 165

\bibitem[{{Lazareff} {et~al.}(2017){Lazareff}, {Berger}, {Kluska}, {Le
  Bouquin}, {Benisty}, {Malbet}, {Koen}, {Pinte}, {Thi}, {Absil}, {Baron},
  {Delboulb{\'e}}, {Duvert}, {Isella}, {Jocou}, {Juhasz}, {Kraus}, {Lachaume},
  {M{\'e}nard}, {Millan-Gabet}, {Monnier}, {Moulin}, {Perraut}, {Rochat},
  {Soulez}, {Tallon}, {Thi{\'e}baut}, {Traub}, \& {Zins}}]{laz17}
{Lazareff}, B., {Berger}, J.~P., {Kluska}, J., {et~al.} 2017, \aap, 599, A85

\bibitem[{{Lesur} {et~al.}(2023){Lesur}, {Flock}, {Ercolano}, {Lin}, {Yang},
  {Barranco}, {Benitez-Llambay}, {Goodman}, {Johansen}, {Klahr}, {Laibe},
  {Lyra}, {Marcus}, {Nelson}, {Squire}, {Simon}, {Turner}, {Umurhan}, \&
  {Youdin}}]{les23}
{Lesur}, G., {Flock}, M., {Ercolano}, B., {et~al.} 2023, in Astronomical
  Society of the Pacific Conference Series, Vol. 534, Astronomical Society of
  the Pacific Conference Series, ed. S.~{Inutsuka}, Y.~{Aikawa}, T.~{Muto},
  K.~{Tomida}, \& M.~{Tamura}, 465

\bibitem[{{Lesur}(2021)}]{les21}
{Lesur}, G. R.~J. 2021, \aap, 650, A35

\bibitem[{{Levermore} \& {Pomraning}(1981)}]{lev81}
{Levermore}, C.~D. \& {Pomraning}, G.~C. 1981, \apj, 248, 321

\bibitem[{{Lissauer} {et~al.}(2014){Lissauer}, {Dawson}, \& {Tremaine}}]{lis14}
{Lissauer}, J.~J., {Dawson}, R.~I., \& {Tremaine}, S. 2014, \nat, 513, 336

\bibitem[{{Lissauer} {et~al.}(2011){Lissauer}, {Ragozzine}, {Fabrycky},
  {Steffen}, {Ford}, {Jenkins}, {Shporer}, {Holman}, {Rowe}, {Quintana},
  {Batalha}, {Borucki}, {Bryson}, {Caldwell}, {Carter}, {Ciardi}, {Dunham},
  {Fortney}, {Gautier}, {Howell}, {Koch}, {Latham}, {Marcy}, {Morehead}, \&
  {Sasselov}}]{lis11}
{Lissauer}, J.~J., {Ragozzine}, D., {Fabrycky}, D.~C., {et~al.} 2011, \apjs,
  197, 8

\bibitem[{{Malygin} {et~al.}(2014){Malygin}, {Kuiper}, {Klahr}, {Dullemond}, \&
  {Henning}}]{mal14}
{Malygin}, M.~G., {Kuiper}, R., {Klahr}, H., {Dullemond}, C.~P., \& {Henning},
  T. 2014, \aap, 568, A91

\bibitem[{{Marleau} {et~al.}(2019){Marleau}, {Mordasini}, \& {Kuiper}}]{mar19}
{Marleau}, G.-D., {Mordasini}, C., \& {Kuiper}, R. 2019, \apj, 881, 144

\bibitem[{{Masset} {et~al.}(2006){Masset}, {Morbidelli}, {Crida}, \&
  {Ferreira}}]{mas06}
{Masset}, F.~S., {Morbidelli}, A., {Crida}, A., \& {Ferreira}, J. 2006, \apj,
  642, 478

\bibitem[{{Melon Fuksman} \& {Klahr}(2022)}]{fuk22}
{Melon Fuksman}, J.~D. \& {Klahr}, H. 2022, \apj, 936, 16

\bibitem[{{Mendigut{\'\i}a} {et~al.}(2015){Mendigut{\'\i}a}, {de Wit},
  {Oudmaijer}, {Fairlamb}, {Carciofi}, {Ilee}, \& {Vieira}}]{mend15}
{Mendigut{\'\i}a}, I., {de Wit}, W.~J., {Oudmaijer}, R.~D., {et~al.} 2015,
  \mnras, 453, 2126

\bibitem[{{Menu} {et~al.}(2015){Menu}, {van Boekel}, {Henning}, {Leinert},
  {Waelkens}, \& {Waters}}]{men15}
{Menu}, J., {van Boekel}, R., {Henning}, T., {et~al.} 2015, \aap, 581, A107

\bibitem[{{Millan-Gabet} {et~al.}(2016){Millan-Gabet}, {Che}, {Monnier},
  {Sitko}, {Russell}, {Grady}, {Day}, {Perry}, {Harries}, {Aarnio}, {Colavita},
  {Wizinowich}, {Ragland}, \& {Woillez}}]{mil16}
{Millan-Gabet}, R., {Che}, X., {Monnier}, J.~D., {et~al.} 2016, \apj, 826, 120

\bibitem[{{Mulders} \& {Dominik}(2012)}]{mul12}
{Mulders}, G.~D. \& {Dominik}, C. 2012, \aap, 539, A9

\bibitem[{{Mulders} {et~al.}(2015){Mulders}, {Pascucci}, \& {Apai}}]{mul15}
{Mulders}, G.~D., {Pascucci}, I., \& {Apai}, D. 2015, \apj, 798, 112

\bibitem[{{Mulders} {et~al.}(2018){Mulders}, {Pascucci}, {Apai}, \&
  {Ciesla}}]{mul18}
{Mulders}, G.~D., {Pascucci}, I., {Apai}, D., \& {Ciesla}, F.~J. 2018, \aj,
  156, 24

\bibitem[{{Natta} {et~al.}(2001){Natta}, {Prusti}, {Neri}, {Wooden}, {Grinin},
  \& {Mannings}}]{nat01}
{Natta}, A., {Prusti}, T., {Neri}, R., {et~al.} 2001, \aap, 371, 186

\bibitem[{{Nesvorn{\'y}} {et~al.}(2022){Nesvorn{\'y}}, {Chrenko}, \&
  {Flock}}]{nes22}
{Nesvorn{\'y}}, D., {Chrenko}, O., \& {Flock}, M. 2022, \apj, 925, 38

\bibitem[{{Oka} {et~al.}(2011){Oka}, {Nakamoto}, \& {Ida}}]{oka11}
{Oka}, A., {Nakamoto}, T., \& {Ida}, S. 2011, \apj, 738, 141

\bibitem[{{Petigura} {et~al.}(2018){Petigura}, {Marcy}, {Winn}, {Weiss},
  {Fulton}, {Howard}, {Sinukoff}, {Isaacson}, {Morton}, \& {Johnson}}]{pet18}
{Petigura}, E.~A., {Marcy}, G.~W., {Winn}, J.~N., {et~al.} 2018, \aj, 155, 89

\bibitem[{{Schobert} \& {Peeters}(2021)}]{scho21}
{Schobert}, B.~N. \& {Peeters}, A.~G. 2021, \aap, 651, A27

\bibitem[{{Schobert} {et~al.}(2019){Schobert}, {Peeters}, \& {Rath}}]{scho19}
{Schobert}, B.~N., {Peeters}, A.~G., \& {Rath}, F. 2019, \apj, 881, 56

\bibitem[{{Seok} \& {Li}(2017)}]{seo17}
{Seok}, J.~Y. \& {Li}, A. 2017, \apj, 835, 291

\bibitem[{{Serra D{\'\i}az-Cano} \& {Jones}(2008)}]{ser08}
{Serra D{\'\i}az-Cano}, L. \& {Jones}, A.~P. 2008, \aap, 492, 127

\bibitem[{{Setterholm} {et~al.}(2018){Setterholm}, {Monnier}, {Davies},
  {Kreplin}, {Kraus}, {Baron}, {Aarnio}, {Berger}, {Calvet}, {Cur{\'e}},
  {Kanaan}, {Kloppenborg}, {Le Bouquin}, {Millan-Gabet}, {Rubinstein}, {Sitko},
  {Sturmann}, {ten Brummelaar}, \& {Touhami}}]{Set18}
{Setterholm}, B.~R., {Monnier}, J.~D., {Davies}, C.~L., {et~al.} 2018, \apj,
  869, 164

\bibitem[{{Shakura} \& {Sunyaev}(1973)}]{sha73}
{Shakura}, N.~I. \& {Sunyaev}, R.~A. 1973, \aap, 24, 337

\bibitem[{{Thi} {et~al.}(2019){Thi}, {Lesur}, {Woitke}, {Kamp}, {Rab}, \&
  {Carmona}}]{thi19}
{Thi}, W.~F., {Lesur}, G., {Woitke}, P., {et~al.} 2019, \aap, 632, A44

\bibitem[{{Tschudi} \& {Schmid}(2021)}]{tsc21}
{Tschudi}, C. \& {Schmid}, H.~M. 2021, \aap, 655, A37

\bibitem[{{Turner} {et~al.}(2014){Turner}, {Benisty}, {Dullemond}, \&
  {Hirose}}]{tur14}
{Turner}, N.~J., {Benisty}, M., {Dullemond}, C.~P., \& {Hirose}, S. 2014, \apj,
  780, 42

\bibitem[{{Ueda} {et~al.}(2021){Ueda}, {Flock}, \& {Birnstiel}}]{ued21}
{Ueda}, T., {Flock}, M., \& {Birnstiel}, T. 2021, \apjl, 914, L38

\bibitem[{{Ueda} {et~al.}(2019){Ueda}, {Flock}, \& {Okuzumi}}]{ued19}
{Ueda}, T., {Flock}, M., \& {Okuzumi}, S. 2019, \apj, 871, 10

\bibitem[{{Ueda} {et~al.}(2017){Ueda}, {Okuzumi}, \& {Flock}}]{ued17}
{Ueda}, T., {Okuzumi}, S., \& {Flock}, M. 2017, \apj, 843, 49

\bibitem[{{Varga} {et~al.}(2021){Varga}, {Hogerheijde}, {van Boekel},
  {Klarmann}, {Petrov}, {Waters}, {Lagarde}, {Pantin}, {Berio}, {Weigelt},
  {Robbe-Dubois}, {Lopez}, {Millour}, {Augereau}, {Meheut}, {Meilland},
  {Henning}, {Jaffe}, {Bettonvil}, {Bristow}, {Hofmann}, {Matter}, {Zins},
  {Wolf}, {Allouche}, {Donnan}, {Schertl}, {Dominik}, {Heininger}, {Lehmitz},
  {Cruzal{\`e}bes}, {Glindemann}, {Meisenheimer}, {Paladini}, {Sch{\"o}ller},
  {Woillez}, {Venema}, {Kokoulina}, {Yoffe}, {{\'A}brah{\'a}m}, {Abadie},
  {Abuter}, {Accardo}, {Adler}, {Ag{\'o}cs}, {Antonelli}, {B{\"o}hm}, {Bailet},
  {Bazin}, {Beckmann}, {Beltran}, {Boland}, {Bourget}, {Brast}, {Bresson},
  {Burtscher}, {Castillo}, {Chelli}, {Cid}, {Clausse}, {Connot}, {Conzelmann},
  {Danchi}, {De Haan}, {Delbo}, {Ebert}, {Elswijk}, {Fantei}, {Frahm},
  {G{\'a}mez Rosas}, {Gabasch}, {Gallenne}, {Garces}, {Girard}, {Gont{\'e}},
  {Gonz{\'a}lez Herrera}, {Graser}, {Guajardo}, {Guitton}, {Haubois}, {Hron},
  {Hubin}, {Huerta}, {Isbell}, {Ives}, {Jakob}, {Jask{\'o}}, {Jochum}, {Klein},
  {Kragt}, {Kroes}, {Kuindersma}, {Labadie}, {Laun}, {Le Poole}, {Leinert},
  {Lizon}, {Lopez}, {M{\'e}rand}, {Marcotto}, {Mauclert}, {Maurer}, {Mehrgan},
  {Meisner}, {Meixner}, {Mellein}, {Mohr}, {Morel}, {Mosoni}, {Navarro},
  {Neumann}, {Nu{\ss}baum}, {Pallanca}, {Pasquini}, {Percheron}, {Pott},
  {Pozna}, {Ridinger}, {Rigal}, {Riquelme}, {Rivinius}, {Roelfsema}, {Rohloff},
  {Rousseau}, {Schuhler}, {Schuil}, {Soulain}, {Stee}, {Stephan}, {ter Horst},
  {Tromp}, {Vakili}, {van Duin}, {Vinther}, {Wittkowski}, \& {Wrhel}}]{var21}
{Varga}, J., {Hogerheijde}, M., {van Boekel}, R., {et~al.} 2021, \aap, 647, A56

\bibitem[{{Varga} {et~al.}(2024){Varga}, {Waters}, {Hogerheijde}, {van Boekel},
  {Matter}, {Lopez}, {Perraut}, {Chen}, {Nadella}, {Wolf}, {Dominik},
  {K{\'o}sp{\'a}l}, {{\'A}brah{\'a}m}, {Augereau}, {Boley}, {Bourdarot},
  {Caratti O Garatti}, {Cruz-S{\'a}enz de Miera}, {Danchi}, {G{\'a}mez Rosas},
  {Henning}, {Hofmann}, {Houll{\'e}}, {Isbell}, {Jaffe}, {Juh{\'a}sz},
  {Kecskem{\'e}thy}, {Kobus}, {Kokoulina}, {Labadie}, {Lykou}, {Millour},
  {Mo{\'o}r}, {Moruj{\~a}o}, {Pantin}, {Schertl}, {Scheuck}, {van Haastere},
  {Weigelt}, {Woillez}, {Woitke}, {Matisse Collaboration}, \& {Gravity
  Collaboration}}]{var24}
{Varga}, J., {Waters}, L.~B.~F.~M., {Hogerheijde}, M., {et~al.} 2024, \aap,
  681, A47

\bibitem[{{Vinkovi{\'c}}(2012)}]{vin12}
{Vinkovi{\'c}}, D. 2012, \mnras, 420, 1541

\bibitem[{{Vinkovi{\'c}} {et~al.}(2006){Vinkovi{\'c}}, {Ivezi{\'c}},
  {Jurki{\'c}}, \& {Elitzur}}]{vin06}
{Vinkovi{\'c}}, D., {Ivezi{\'c}}, {\v Z}., {Jurki{\'c}}, T., \& {Elitzur}, M.
  2006, \apj, 636, 348

\bibitem[{{Vinkovi{\'c}} \& {{\v{C}}emelji{\'c}}(2021)}]{vin21}
{Vinkovi{\'c}}, D. \& {{\v{C}}emelji{\'c}}, M. 2021, \mnras, 500, 506

\bibitem[{{Weiss} {et~al.}(2023){Weiss}, {Millholland}, {Petigura}, {Adams},
  {Batygin}, {Block}, \& {Mordasini}}]{Wei23}
{Weiss}, L.~M., {Millholland}, S.~C., {Petigura}, E.~A., {et~al.} 2023, in
  Astronomical Society of the Pacific Conference Series, Vol. 534, Astronomical
  Society of the Pacific Conference Series, ed. S.~{Inutsuka}, Y.~{Aikawa},
  T.~{Muto}, K.~{Tomida}, \& M.~{Tamura}, 863

\bibitem[{{Wichittanakom} {et~al.}(2020){Wichittanakom}, {Oudmaijer},
  {Fairlamb}, {Mendigut{\'\i}a}, {Vioque}, \& {Ababakr}}]{wic20}
{Wichittanakom}, C., {Oudmaijer}, R.~D., {Fairlamb}, J.~R., {et~al.} 2020,
  \mnras, 493, 234

\bibitem[{{Williams} \& {Mohanty}(2025)}]{wil25}
{Williams}, M. \& {Mohanty}, S. 2025, \mnras, 536, 1518

\bibitem[{{Woitke} {et~al.}(2018){Woitke}, {Helling}, {Hunter}, {Millard},
  {Turner}, {Worters}, {Blecic}, \& {Stock}}]{woi18}
{Woitke}, P., {Helling}, C., {Hunter}, G.~H., {et~al.} 2018, \aap, 614, A1

\bibitem[{{Woitke} {et~al.}(2016){Woitke}, {Min}, {Pinte}, {Thi}, {Kamp},
  {Rab}, {Anthonioz}, {Antonellini}, {Baldovin-Saavedra}, {Carmona}, {Dominik},
  {Dionatos}, {Greaves}, {G{\"u}del}, {Ilee}, {Liebhart}, {M{\'e}nard},
  {Rigon}, {Waters}, {Aresu}, {Meijerink}, \& {Spaans}}]{woi16}
{Woitke}, P., {Min}, M., {Pinte}, C., {et~al.} 2016, \aap, 586, A103

\bibitem[{{Wu} \& {Lithwick}(2021)}]{wu21}
{Wu}, Y. \& {Lithwick}, Y. 2021, \apj, 923, 123

\end{thebibliography}

\newpage

\begin{appendix}
\onecolumn
\section{Equations for the radiation magnetostatic case}
\label{apa}
In this section, we display the magnetohydrostatic equations, which are solved to derive the azimuthal velocity and the density structure. Using the corresponding divergence operator, we can determine $v_\phi^2$  with
\begin{equation}
v_\mathbf{\phi}^2 =  r \left( \frac{\partial \Phi}{\partial r} + \frac{\partial P}{\partial r} \right )  +  \frac{1}{\rho} \left ( B_\mathbf{\phi}^2 + B_\mathbf{\theta}^2 - 2 B_r^2 - \frac{B_\mathbf{r} B_\mathbf{\theta}}{\tan{\theta}}  \right  )   - \frac{B_\mathbf{r}}{\rho} \left(r \frac{\partial B_\mathbf{r}}{\partial r} + \frac{\partial B_\mathbf{\theta}}{\partial \theta}  \right ) + \frac{B_\mathbf{\theta}}{\rho} \left (r \frac{\partial B_\mathbf{\theta}}{\partial r} - \frac{\partial B_\mathbf{r}}{\partial \theta}  \right )  + \frac{B_\mathbf{\phi}}{\rho} r \frac{\partial B_\mathbf{\phi}}{\partial r}
\end{equation}

and then do a half-cell integration to obtain the new density using
\begin{equation}
\frac{1}{\rho} \frac{\partial \rho}{\partial \theta} =  \frac{1}{T^{\mathbf{n}}} \left ( \frac{v_\mathbf{\phi}^2}{\tan{\theta}} - \frac{\partial T^\mathbf{n}}{\partial \theta} - \frac{\partial \Phi}{\partial \theta} \right ) + \frac{1}{\rho T^\mathbf{n}} \left ( 3 B_\mathbf{\theta} B_\mathbf{r} + \frac{B_\mathbf{\theta}^2 - B_\mathbf{\phi}^2}{\tan{\theta}} \right ) + \frac{B_\mathbf{r}}{\rho T^\mathbf{n}} \left ( r \frac{\partial B_\mathbf{\theta}}{\partial r} - \frac{\partial B_\mathbf{r}}{\partial \theta}  \right ) +  \frac{B_\mathbf{\theta}}{\rho T^\mathbf{n}} \left ( r \frac{\partial B_\mathbf{r}}{\partial r} + \frac{\partial B_\mathbf{\theta}}{\partial \theta}\right ) - \frac{B_\mathbf{\phi}}{\rho T^\mathbf{n}} \frac{\partial B_\mathbf{\phi}}{\partial \theta}.
\end{equation}
The magnetic field remains fixed during all iteration steps. 

\section{\texttt{GGchem} and \texttt{Optool}}
\label{apb}

\begin{figure}[h!]
    \centering
\begin{subfigure}[b]{.475\linewidth}
    \includegraphics[width=\linewidth]{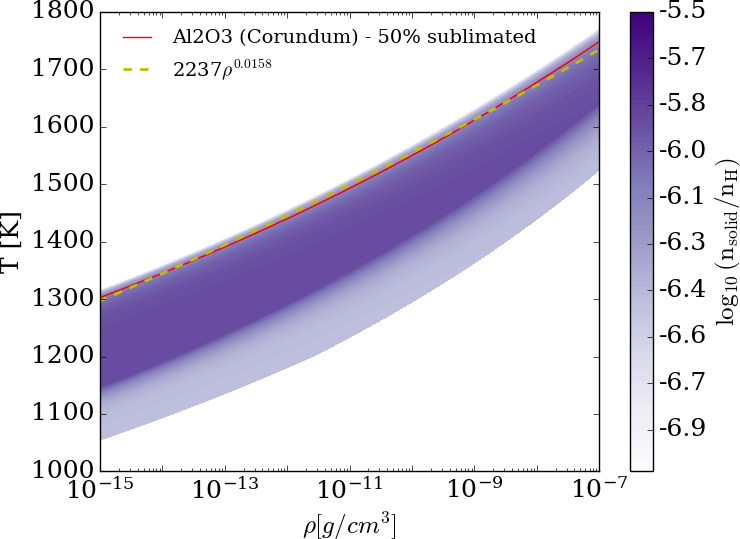}
        \caption{Corundum}
    \label{fig:ggchem_cor}
\end{subfigure}
\hfill
\begin{subfigure}[b]{.475\linewidth}
    \includegraphics[width=\linewidth]{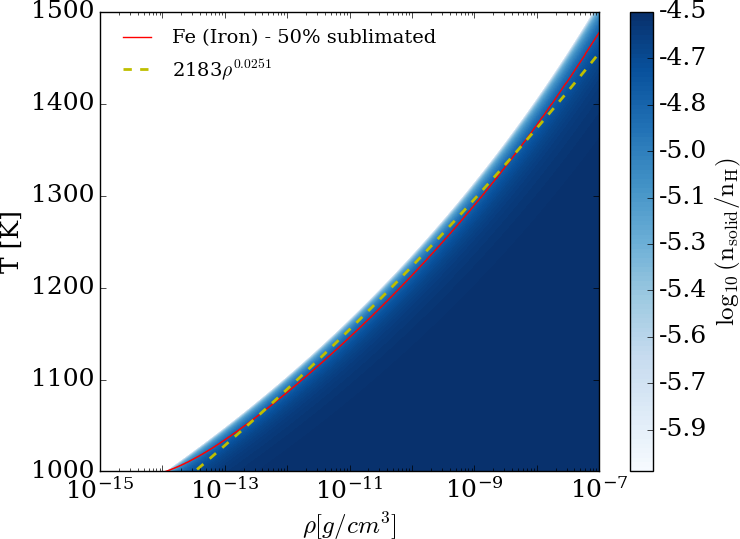}
    \caption{Iron}
    \label{fig:ggchem_fe}
\end{subfigure}
\medskip
\centering
\begin{subfigure}[b]{.475\linewidth}
    \includegraphics[width=\linewidth]{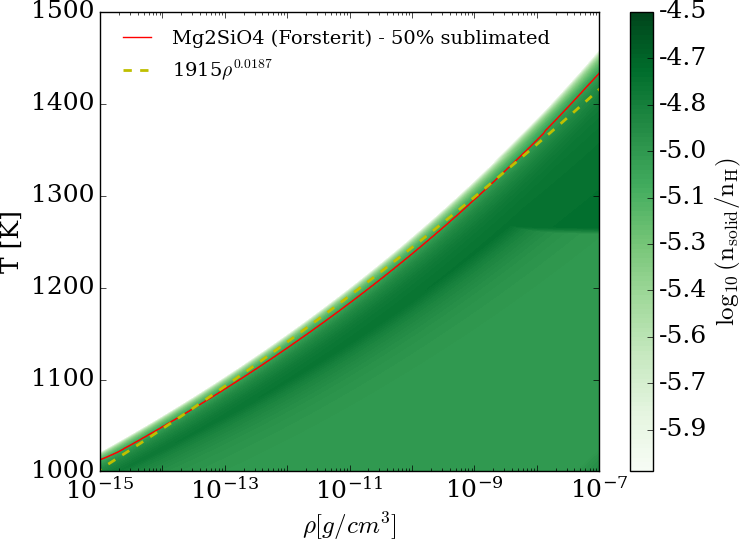}
    \caption{Forsterite}
    \label{fig:ggchem_fo}
\end{subfigure}
\hfill
\begin{subfigure}[b]{.475\linewidth}
    \includegraphics[width=\linewidth]{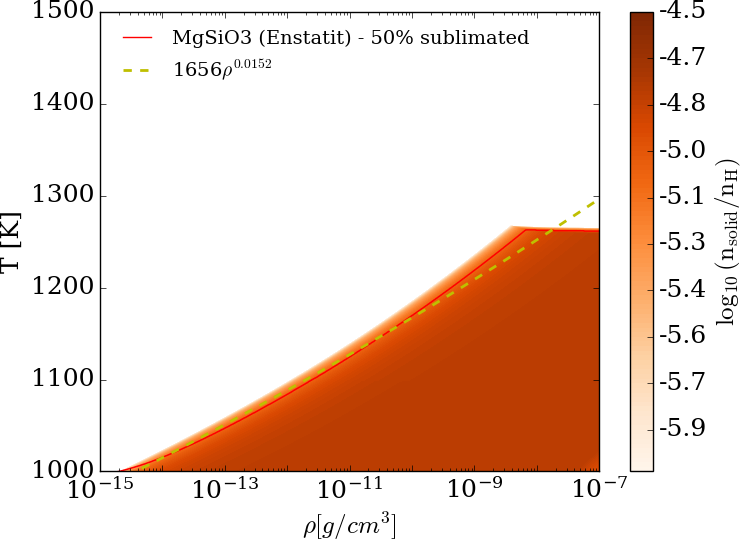}
    \caption{Enstatite}
    \label{fig:ggchem_en}
\end{subfigure}

\caption{Dust species abundance in the density and temperature plane calculated with \texttt{GGchem}.}
\label{fig:ggchem}
   \end{figure}

\begin{figure}
\includegraphics[width=0.49\textwidth]{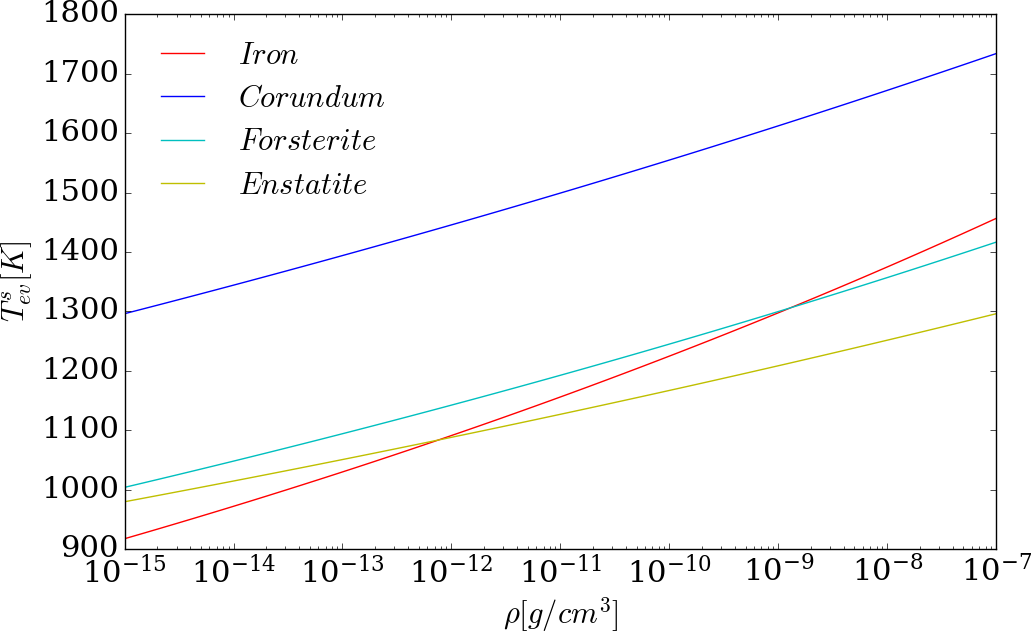}
\includegraphics[width=0.49\textwidth]{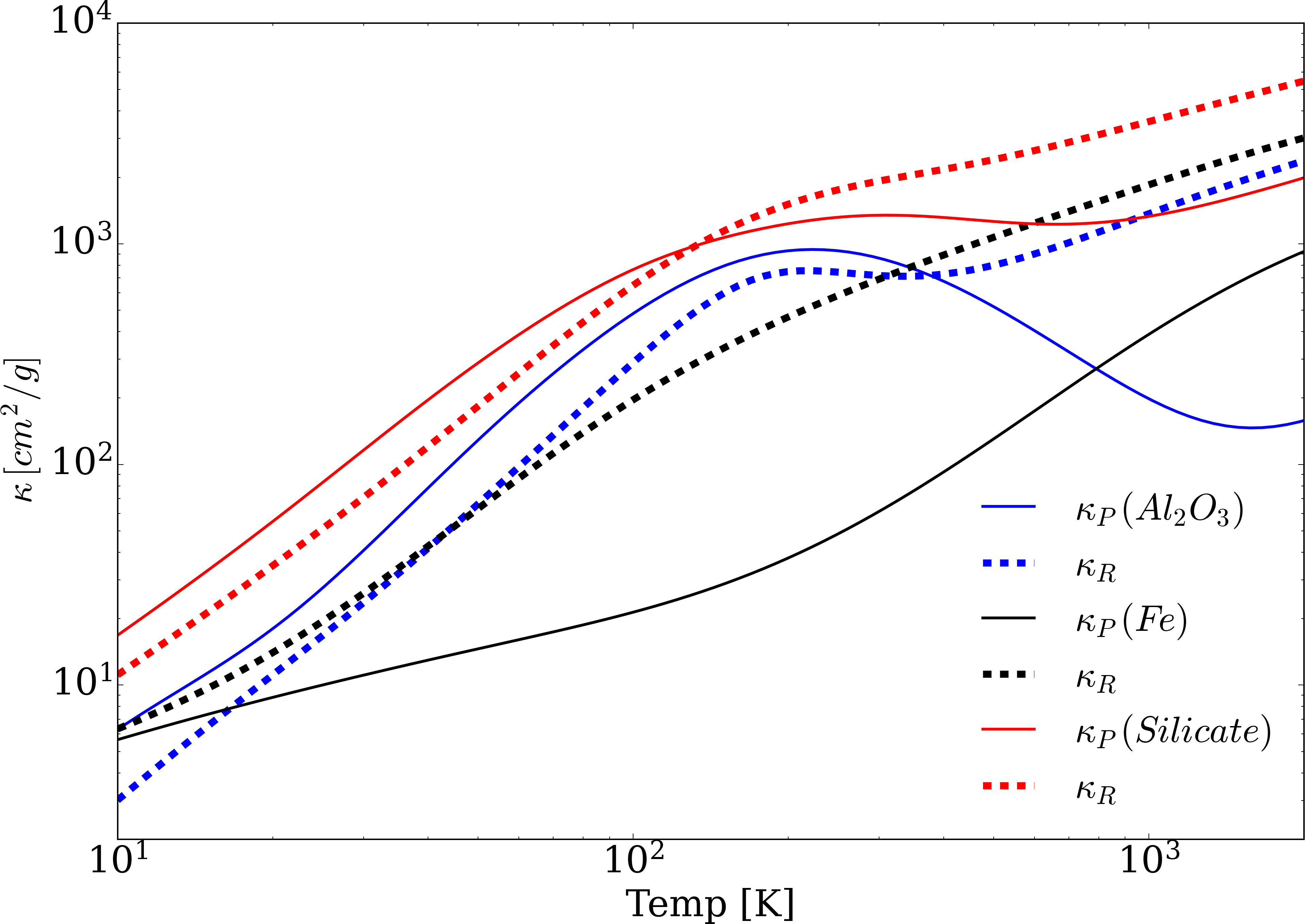}
\caption{Left: Derived fitting functions of the sublimation temperature for the dust species. Right: Planck and Rosseland opacities dependent on temperature for the three groups of dust species. The Rosseland opacities were calculated using the total extinction opacity.} 
\label{fig:tsub}
\end{figure}

For \texttt{GGchem}, we use the standard model using solar abundances, with equilibrium condensation and 1800 points in each direction of the 2D temperature and density regime \citep{woi18}. We show the abundance of the different dust species derived in Fig.~\ref{fig:ggchem}. Each of the four subplots shows the abundance of corundum, iron, forsterite, and enstatite for various local gas densities and temperatures. The red lines show where 50\% of the maximum abundance is reached. The dashed lines show the fitting function we derived and used for our models. The fitting functions are summarized in Table~\ref{tab:dust_spec} and Fig.~\ref{fig:tsub}, left. The command to derive the corresponding dust opacities is summarized in Table~\ref{tab:optool}. The first command calculates the opacity for the two silicate dust grain groups. We follow the standard DIANA opacities \citep{woi16}, which assume a specific pyroxene (70\% Mg) and carbon in a mass ratio $0.87/0.13$ and with a porosity of 25\%. We assume a grain size distribution with grains between $\rm 0.05 \mu$m and $\rm 10 \mu$m with the typical grain size power law of $a^{-3.5}$. In our models, the DIANA opacities represent the opacity of the enstatite and forsterite, the two most significant mass fractions of the dust mass composition; see Table~\ref{tab:comp}. The opacity for the corundum and iron are calculated using the same dust size distribution with a zero porosity level. Finally, we note that further improvements of the opacity are possible as the DIANA opacity is based on Mg0.7Fe0.3SiO$_3$ plus carbon material, while forsterite (Mg$_2$SiO$_4$) and enstatite (MgSiO$_3$) are purely composed of Mg, Si, and O. 
The temperature-dependent Rosseland and Planck mean opacities used for the model are shown in Fig.~\ref{fig:tsub}, right. We also note that we assume three separate dust species when averaging the mean opacity in section~\ref{sec:op} instead of forming a new complex grain species with multiple ingredients. By calculating with Optool for a grain mixture of Corundum, Iron, and Silicates, each with 33\% mass fraction, we found an absorption opacity at 1 $\mu m$ larger by around 67\% compared to just averaging over the grain opacity of the three species with pure material composition.
Determining the opacity in each cell for a specific mixture of dust species would require running \texttt{Optool} for each dust mixture in every cell, which would go beyond the scope of this work.

\begin{table}
    \centering
    \caption{Commands to generate the opacity using \texttt{Optool} \label{tab:optool} }
    \begin{tabular}{l|l}
    Dust species & \texttt{Optool} command\\ 
    \hline
      Silicate &  optool pyr-mg70 0.87 c 0.13 -p 0.25 -radmc \\
        Corundum & optool cor -a 0.05 10 -radmc\\
         Iron      & optool -c fe-c -a 0.05 10 -radmc   \\
    \end{tabular}
\end{table}

\section{Pure silicate grains test}
\label{apc}
We explain here the model \texttt{M001\_S} and \texttt{M001\_S\_TS}, which both have only one dust species. The total dust mass stays the same. However, we had to replace Eq.~\ref{eq:d2g} with the previous formula used by \citet{flo19} to resolve the optical depth transition due to the steep rise in dust density. The other parameters and setup remain the same. The rim has a steeper and more considerable vertical extent, similar to that found in \citep{flo16}. Further, for model \texttt{M001\_S\_TS}, we fix the evaporation temperature to a constant temperature value of 1550~K without any density dependence. This leads to an even steeper wall, similar to that described in \citet{chr24}.

\section{Resolution test}
\label{apd}
\begin{figure*}[h!]
\includegraphics[width=0.49\textwidth]{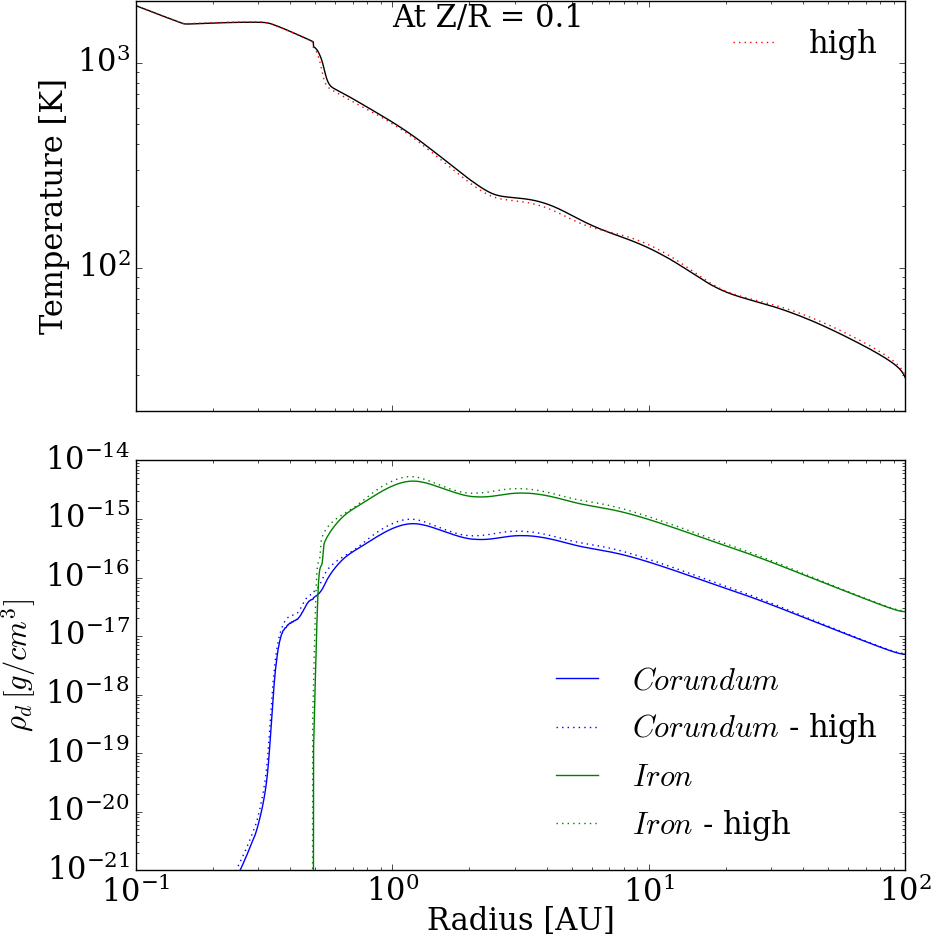}
\includegraphics[width=0.49\textwidth]{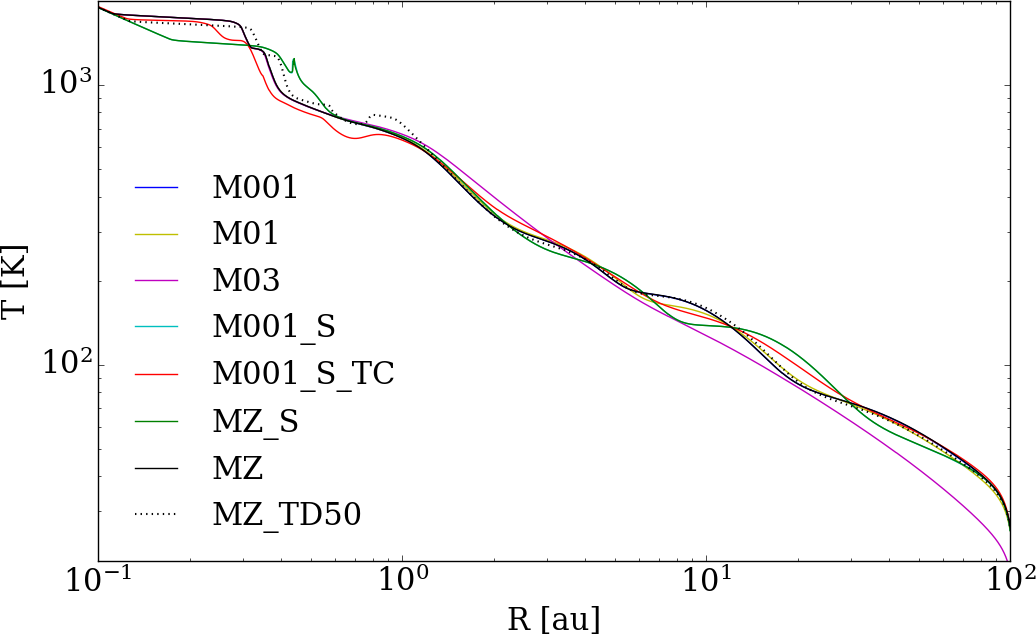}
\caption{Convergence test for model \texttt{M001} with double resolution. The profiles of temperature (top left) and dust densities (bottom left) remain very similar. Right: 1D midplane temperature profile for all models listed in Table~\ref{tab:info}.} 
\label{fig:comp_high}
\end{figure*}

In this section, we test the convergence of the models. We especially want to test the new sublimation function, see Eq.~\ref{eq:d2g}.  We double the resolution of model \texttt{M001}, having 4608 cells in the radial direction and 432 in the $\theta$ direction. We compare the radial profile of the temperature and the density profiles of corundum and iron at the height of $Z/R =0.1$ in Fig.~\ref{fig:comp_high}, left. The results are very similar and show convergence. An overview of the temperature profile from all models in Table~\ref{tab:info} is shown in Fig.~\ref{fig:comp_high}, right.

\FloatBarrier

\section{Strongly magnetized models}
\label{ape}
\begin{figure}
\resizebox{\hsize}{!}{\includegraphics{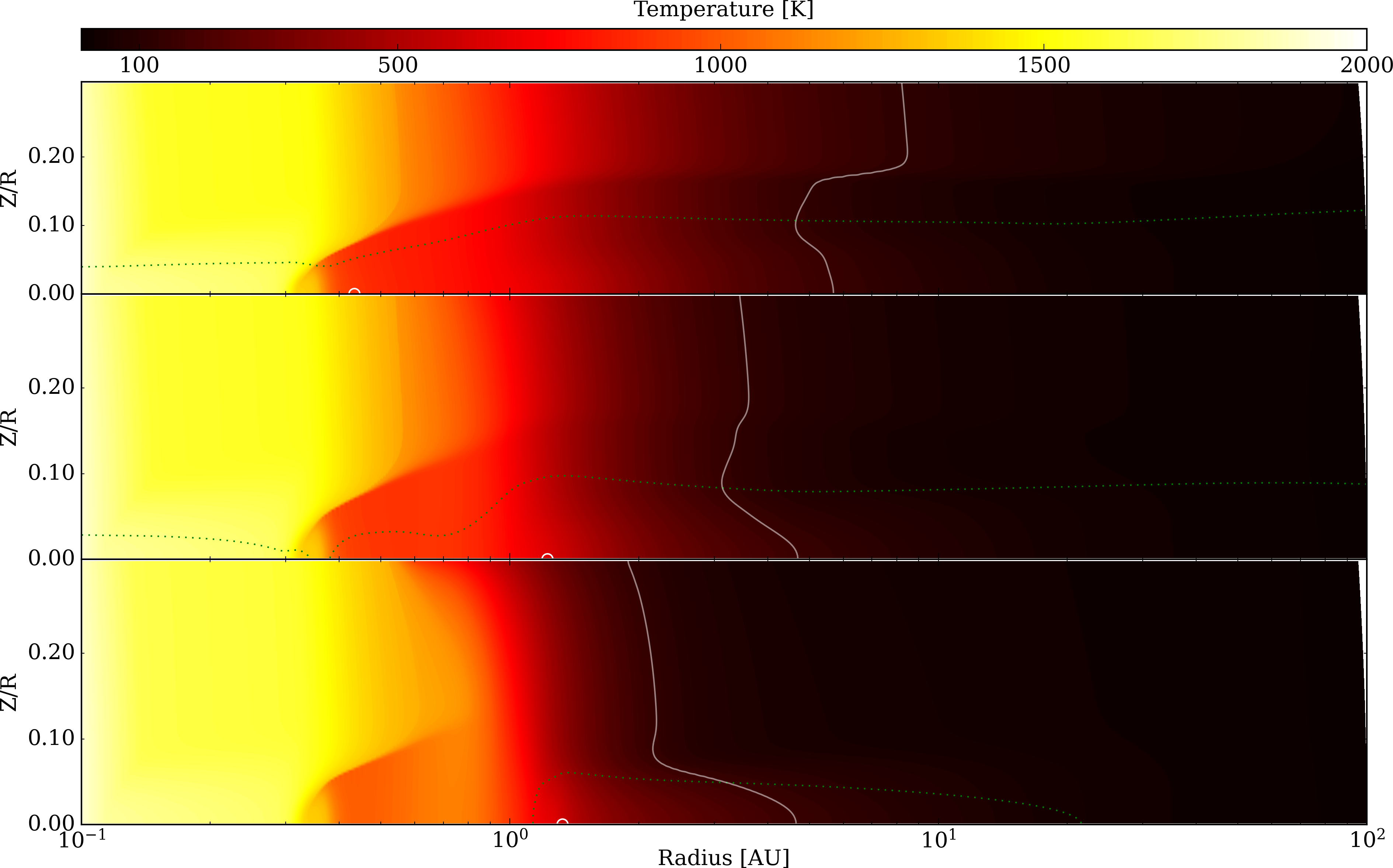}}
\caption{2D temperature profile with increasing magnetization from top to bottom showing models: \texttt{M06},\texttt{M1} and \texttt{M3}. The green dotted line shows the plasma beta unity line.}  
\label{fig:tempmag_part2}
\end{figure}

\begin{figure}
\includegraphics[width=0.49\textwidth]{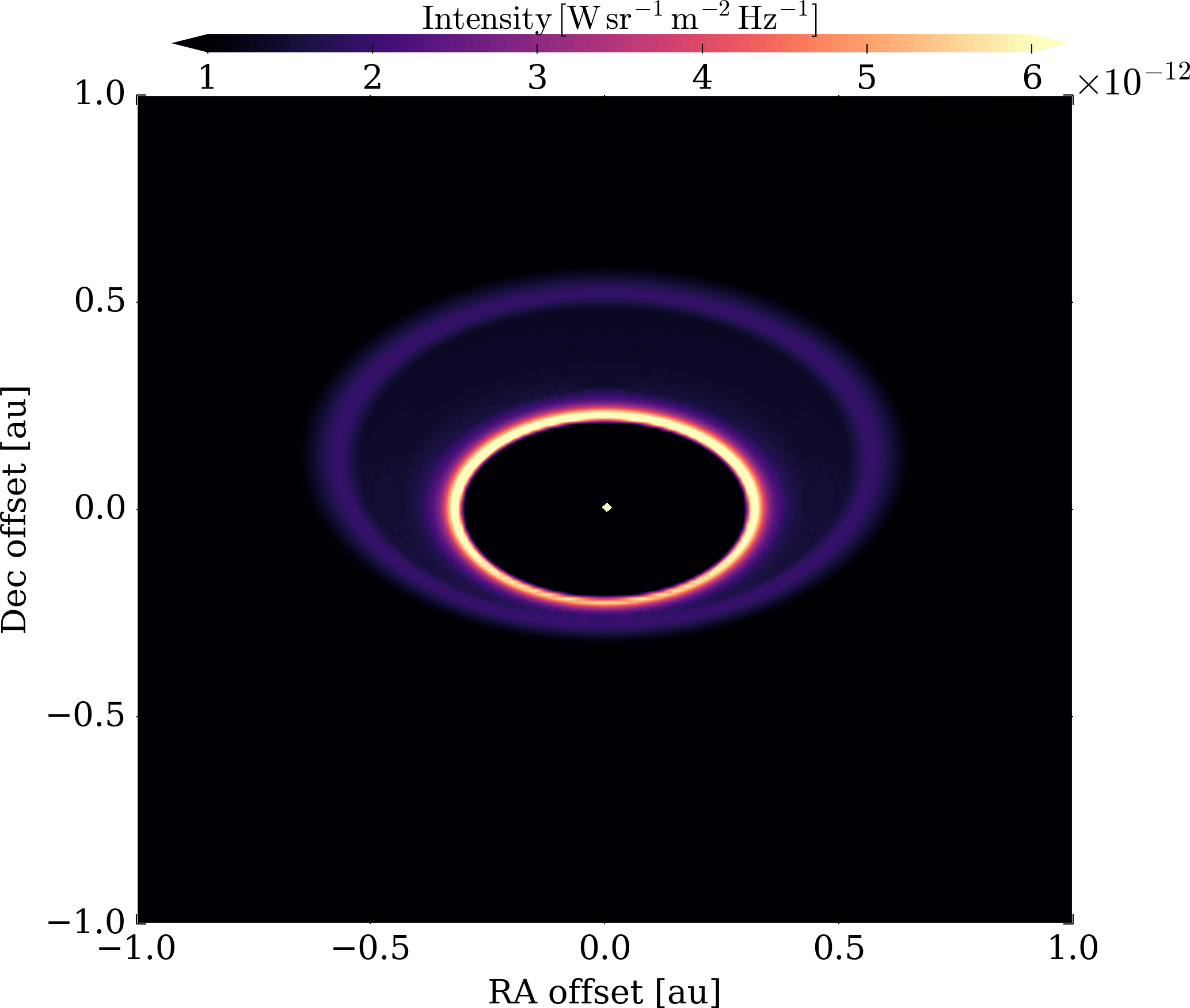}
\includegraphics[width=0.49\textwidth]{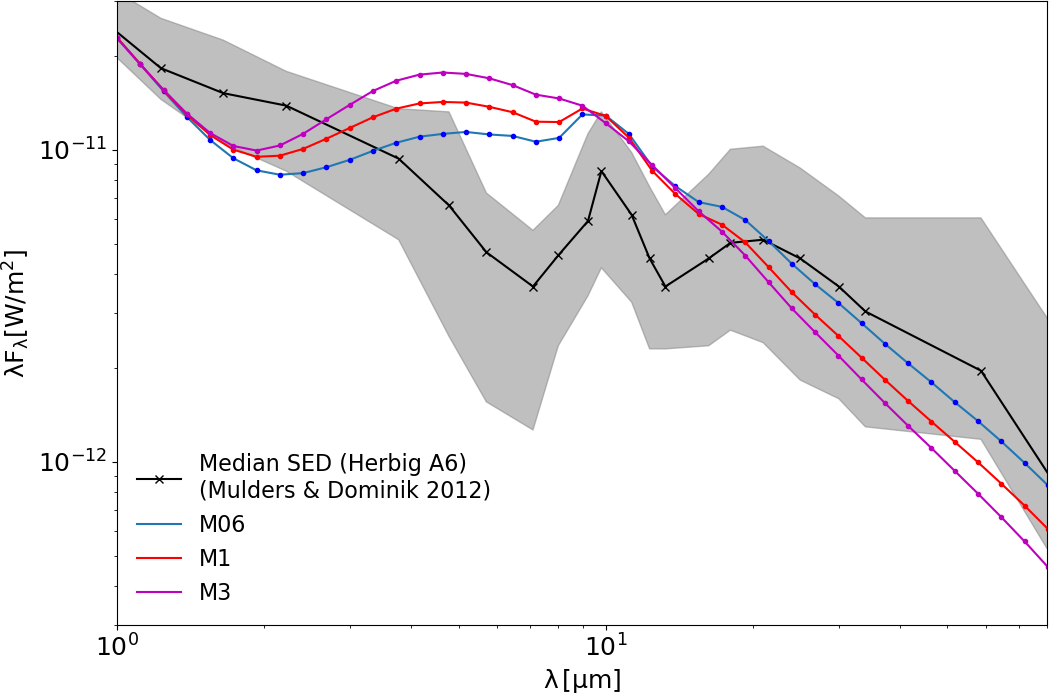}
\caption{Left: 2D intensity map calculated for an inclination of 45$^\circ$ at 2.2~$\mu m$ zoomed into the 1~au region shown for the very magnetized model \texttt{M3}. Right: SED for the high magnetized models from \texttt{M06} to \texttt{M3} showing the 64 calculated wavelength points (dots). The grey shaded color and black solid line show the median SED constructed by \citet{mul12}. All values were normalized to a distance at 100~pc.} 
\label{fig:rad_int_2drim_m3}
\end{figure}

Fig.~\ref{fig:tempmag_part2} shows the temperature profiles for the models with very high magnetization; see Table~\ref{tab:model_mag}. Looking now at models \texttt{M06} and \texttt{M1} (first and second row), the ice-line continues to move inward due to the steeper temperature gradient caused by the higher density in the disk atmosphere. The disk becomes more similar to an optically thick envelope. By further increasing the magnetization in model \texttt{M1} and \texttt{M3}, the disk becomes fully magnetically dominated with the plasma beta below unity inside 1.2~au. In the optical thick envelope, the stellar radiation is quickly absorbed. In model \texttt{M3}, the local density becomes so high that it becomes optically thick even close to the vertical boundary. Close to the vertical boundary, the temperature is affected by the boundary condition. The dust trap moves first inward, reaching 0.43 au for model \texttt{M06} and then moves outward beyond 1 au for model \texttt{M1} and \texttt{M3} due to the combined effect of magnetic pressure and the drop in $\alpha$. The magnetic-dominated atmosphere puffs the disk up, leading to a plateau and secondary ring in the K-band image; see Fig.~\ref{fig:rad_int_2drim_m3}, left. The SED shows that the magnetized models show a strong excess at mid-infrared wavelength, which is not seen in the median SED; see Fig.~\ref{fig:rad_int_2drim_m3}, right. An apparent excess in emission between 3 to 10 \mum \ is due to the modified density and temperature structure. Such an excess in mid-infrared was also reported in the work by \citet{tur14}; see their Fig.~7. We obtained a slight increase in H and K-band emission for the magnetized models; however, the overall shape of the SED does not match. As emphasized in the discussion section, the high-magnetized models have a limitation. Magnetic torques in the upper disk layers would lead to a strong accretion rate and, therefore, reduce the surface density substantially, so the overall structure in temperature and density would also change.

\begin{table}
 \centering
 \caption{Overview of the strongly magnetized disk models \label{tab:model_mag}}
\begin{tabular}{lll}
Model & $B_z$ field at 1au [Gauss] & $\beta_z$ at 1au [midplane] \\
\hline
\texttt{M06} & 0.6 & $215.2$\\
\texttt{M1} & 1.0 & $134.2$\\
\texttt{M3} & 3.0 & $0.32$\\
\hline
\end{tabular}
 \tablefoot{From left to right listing: model name, the vertical magnetic field strength at 1au, and the plasma beta of the vertical magnetic field at the midplane at 1au.}
\end{table}

\end{appendix}

\end{document}